\documentclass[conference, compsoc, 10pt, times]{IEEEtran}

\usepackage{cite}
\usepackage{amsmath,amssymb,amsfonts}
\usepackage{algorithmicx}
\usepackage{algpseudocode}
\usepackage{graphicx}
\usepackage{textcomp}
\usepackage{xcolor}
\usepackage{booktabs}
\usepackage{tabularx}

\usepackage[utf8]{inputenc}

\usepackage[hidelinks]{hyperref}
\usepackage{xurl}

\usepackage{multirow}

\usepackage{tikz}
\usepackage{comment}
\graphicspath{{./images/}}

\usepackage{custom_macros}
\usepackage{balance}

\begin{document}

\newlength{\textfloatsepsave}
\setlength{\textfloatsepsave}{\textfloatsep}

\title{Deep perceptual hashing algorithms with hidden dual purpose: \\when client-side scanning does facial recognition}

\author{
\rm{
    Shubham Jain,
    Ana-Maria Cre\c{t}u,
    Antoine Cully,
    and Yves-Alexandre de Montjoye\textsuperscript{\textsection}} 
\\  Imperial College London
}

\maketitle
\begingroup\renewcommand\thefootnote{\textsection}
\footnotetext{Corresponding author at \href{mailto:deMontjoye@imperial.ac.uk}{deMontjoye@imperial.ac.uk}}
\endgroup

\begin{abstract}

End-to-end encryption (E2EE) provides strong technical protections to individuals from interferences. Governments and law enforcement agencies around the world have however raised concerns that E2EE also allows illegal content to be shared undetected. Client-side scanning (CSS), using perceptual hashing (PH) to detect known illegal content before it is shared, is seen as a promising solution to prevent the diffusion of illegal content while preserving encryption. While these proposals raise strong privacy concerns, proponents of the solutions have argued that the risk is limited as the technology has a limited scope: detecting known illegal content.
In this paper, we show that modern perceptual hashing algorithms are actually fairly flexible pieces of technology and that this flexibility could be used by an adversary to add a secondary hidden feature to a client-side scanning system.
More specifically, we show that an adversary providing the PH algorithm can ``hide" a secondary purpose of face recognition of a target individual alongside its primary purpose of image copy detection. 
We first propose a procedure to train a dual-purpose deep perceptual hashing model by jointly optimizing for both the image copy detection and the targeted facial recognition task. Second, we extensively evaluate our dual-purpose model and show it to be able to reliably identify a target individual 67\% of the time while not impacting its performance at detecting illegal content. We also show that our model is neither a general face detection nor a facial recognition model, allowing its secondary purpose to be hidden.
Finally, we show that the secondary purpose can be enabled by adding a single illegal looking image to the database.
Taken together, our results raise concerns that a  deep perceptual hashing-based CSS system could turn billions of user devices into tools to locate targeted individuals.
\end{abstract}

\section{Introduction}

Some of the most commonly used communication and data sharing platforms, ranging from WhatsApp to Signal and iCloud~\cite{whatsappe2ee, signal, icloud} are protected by end-to-end encryption (E2EE). Together, these platforms are used by more than 2B users to privately exchange more than 100B messages and 4.5B images daily~\cite{whatsapp100billion, whatsapp2billion, whatsappblog2017}.
E2EE provides a strong protection to these exchanges. The content of the messages and images shared is indeed not accessible to anyone else but the sender and the intended recipient(s), including the platform providers themselves. These strong technical protections have been essential to create safe and private spaces for personal development, allowing individuals to communicate without interference and providing protection from criminals~\cite{pie2ee}.

Governments and intelligence agencies around the world have however raised concerns that encryption allows criminals to evade detection~\cite{international2020e2ee}. Recently, concerns have focused on how encryption enables illegal content, and more specifically child sexual abuse material (CSAM), to be shared undetected~\cite{ncmec2019}. While a range of potential solutions have been proposed, client-side scanning systems (CSS), have recently been seen as a compromise to detect illegal content while maintaining the privacy of communication. Such systems would indeed flag known illegal images directly on devices, before they are encrypted~\cite{applecsam, kulshrestha2021identifying, levy2022thoughts}, and share them unencrypted with an authority. 

The proposed client-side scanning solutions use perceptual hashing to detect and flag copies of known illegal images. Perceptual hashing (PH) algorithms aim to irreversibly project high-dimensional images to a lower dimensional space, such that images similar to one another are projected close to one another in the lower dimensional space. These are designed to be robust to small modifications in the image, e.g., change of format, rotation.
While designs and deployment details vary, the general idea of perceptual hashing-based CSS proposals is that every image shared by a user would be matched against a database of hashes of known illegal images, and reported if flagged as a match.

Government proposals to mandate the deployment and use of such CSS systems are being discussed around the world, e.g. in the EU~\cite{euregulationcsam} and in the UK~\cite{ukonlineharms}. If adopted, these proposals would provide government agencies with powers to mandate the installation of CSS systems on people's devices. While this raises privacy concerns, proponents have emphasized that CSS is a very specific system with limited capabilities: detecting whether an image being sent is a copy, duplicate or edited, of known illegal content~\cite{ofcomoverview}.

State-of-the-art PH algorithms, including NeuralHash~\cite{applecsam} the algorithm used by Apple, however are large-scale machine learning models whose embeddings are then hashed. These models often contain millions of parameters and are trained on large-scale datasets for detecting images that are approximate copies of one another.
The expressivity of these models has been used against them, e.g. for privacy attacks such as membership inference but also to embed triggers that e.g. lead a model to misclassify the image~\cite{saha2020hidden, li2022backdoor}.

CSS, if mandated, will be installed on billions of devices to scan content before it is encrypted. Law enforcement agencies around the world have long sought to access this data through backdoors, e.g. key-escrow systems~\cite{principles2018key} but also going as far as to add an intentional flaw in a cryptographic standard~\cite{rsa2014}, making it realistic to assume similar attempts could be made when mandated CSS systems are deployed.
This raises concerns that governmental agencies, including intelligence and law enforcement agencies, could embed hidden features in the mandated deep perceptual hashing algorithm used for client-side scanning~\cite{cryptowars, blaze2011key}.

\textbf{Contributions}. In this paper, we show how an adversary could leverage a deep perceptual hashing-based client-side scanning system to scan billions of devices to identify images of a particular target individual. More specifically, we show that the adversary providing the perceptual hashing algorithm, e.g., to E2EE platforms, could ``hide" a secondary purpose of facial recognition of one or more target individual(s) alongside its primary purpose of image copy detection. More specifically, our contributions are: 

\begin{enumerate}
    \item We propose a novel attack model, describing an adversary that provides the perceptual hashing algorithm for a client-side scanning system (CSS) and wants to use this CSS system to find the person(s) of interest (target individual(s)). We propose a methodology to train deep perceptual hashing models to achieve both the primary purpose of image copy detection for flagging known illegal images and a secondary purpose of facial recognition of the target individual. We train a deep perceptual hashing model to jointly optimize for both of these purposes, while also ensuring that the secondary purpose remains ``hidden".
    \item We perform a comprehensive evaluation to measure the ability of deep perceptual hashing models to do image copy detection and facial recognition of the target individual. We show the dual-purpose model to be able to flag previously unknown pictures of the target individual while maintaining its performance on the primary task of image copy detection. 
    \item We further show that the database of illegal images can be innocuously extended with a single illegal-looking image per target individual, such that the client-side scanning system composed of the modified database and corresponding dual-purpose models can detect the pictures of the target individual.
    \item Finally, we show that despite being able to identify pictures containing faces of the target individual, our dual-purpose algorithm is neither a general face detection algorithm, i.e. flagging every image containing a face, nor a general facial recognition algorithm, i.e. matching faces of non-target individuals. Taken together, our results show that deep perceptual hashing-based client-side scanning systems could be built with a hidden purpose that would enable billions of user devices to be used as a surveillance tool.
\end{enumerate}

The rest of the paper is organized as follows: 
Section~\ref{sec:background} gives an overview of client-side scanning and deep perceptual hashing algorithms. 
Section~\ref{sec:attack-model} explains our attack model, while Section~\ref{sec:methodology} details the methodology we use to train single-purpose and dual-purpose models. 
Section~\ref{sec:evaluation} lists the performance metrics we use, and our validation and model selection strategy.
Section~\ref{sec:datasets} provides a detailed description of datasets we use to train, validate, and test our models.
Section~\ref{sec:results} provides the primary results of our experiments. Section~\ref{sec:discussion} discusses the impact of our results and the ``hidden" nature of our dual-purpose CSS system.
Section~\ref{sec:related-work} provides an overview of the related work.

\section{Perceptual hashing-based client-side scanning}\label{sec:background}

\textbf{Client-side scanning (CSS)} is a technology developed to detect illegal content on-device before the content is shared, encrypted, e.g. on WhatsApp, iCloud, Signal~\cite{applecsam}. 
CSS uses perceptual hashing (PH) to detect illegal content by flagging copies, either a duplicate or an edited version, of known illegal images.
A PH algorithm $H$ would be deployed on the user device along with a database $D = \{ d_1, \ldots, d_n \}$ containing hashes $d_i = H(r_i), i = 1 \ldots, n$ of known illegal images $R=\{r_1, \ldots, r_n\}$. 
This allows illegal content to be flagged without having to share a database of illegal images. The database $D$ can furthermore be protected using mechanisms such as cuckoo tables, as in Apple's PH-CSS proposal~\cite{applecsam}. 
This feature, which requires only hashes to be stored on the user device, is a crucial element of CSS, as the storing and sharing of illegal content are criminal offenses according to law in many countries~\cite{ukcsealaw, uscsealaw}.

\textbf{Perceptual hashing (PH) algorithms} aim to irreversibly project high-dimensional images to a lower dimensional space, such that images similar to one another are projected close to one another in the lower dimensional space.
Thus, to retrieve images, from a large database of images, which are similar to a given a query image, one can simply compare the projections generated using PH algorithm. This makes PH algorithms an ideal choice to solve the problem of image copy detection, i.e. finding copies of a given image in a large database~\cite{biswas2021state}.
In the image copy detection task, two images are considered copies of one another if they are exact duplicates or if one of the images is an edited version of the other image.
In CSS, similarly to image copy detection, PH allows an on-device system to flag images that are copies of images in the database $R$ of known illegal images.
PH algorithms like Microsoft's PhotoDNA~\cite{photodna} and Facebook's PDQ~\cite{facebookpdq} have been used in the past to detect known illegal content, albeit in a server-side scanning setup. 
Formally, a perceptual hashing algorithm $H$ maps an input image $X$ to a vector $E$ of length $l$, referred to as \textit{hash}. 
The hash consists of either binary or continuous values. 
We refer the interested reader to Jain et al.~\cite{jain2022adversarial} for a more formal definition of perceptual hashing algorithms.

In CSS, the PH algorithm $H$ is used with a threshold $T$ to detect if an image $X$ is a copy of image in the database $R$.
More specifically, a \textit{query image} $X$ is flagged by CSS if there exists a hash in the database $d_i \in D$ such that $f(H(X), d_i) < T$. 
Here, $f$ denotes a distance function used to measure the dissimilarity between the images, i.e., small values of $f$ indicate more similar images. 
The most popular distance functions to compare hashes are the Hamming distance for binary hashes and the Euclidean distance for continuous hashes.
The threshold $T$ controls the trade-off between recall, i.e. the percentage of true image copy pairs being detected, and precision, i.e. the percentage of detected pairs of images that are actually true image copy pairs.

Modern perceptual hashing algorithms are effectively a trained machine learning model that creates an embedding for each image~\cite{biswas2021state}.
The embeddings could be used directly as the hash or can be converted to a bit-valued hash using locality-sensitive hashing~\cite{applecsam}.
Such models are predominantly built using convolutional neural network (CNN) architectures~\cite{babenko2014neural, tolias2015particular}, and trained on large-scale datasets using self-supervised learning and augmentations~\cite{yokoo2021contrastive, pizzi2022self}. 

\textbf{Micro-Average Precision (${\mu}AP$)} is a standard metric used to measure the quality of a perceptual hashing algorithm $H$ independently of a specific threshold $T$. Similarly to the Area Under the Curve metric, $\mu{AP}$ provides a single number used to compare the ability of PH algorithms to reliably distinguish the pairs of images which are copies of each other from the pairs that contain images which are visually distinct from each other~\cite{douze20212021}. 
$\mu{AP}$ is evaluated by sending a set of query images $Q = \{q_1, q_2, ..., q_m\}$ through the PH algorithm $H$ and matching it against a database of reference images $R$. 
Some images in $Q$ have a valid match in $R$ while some do not. 
The distances $f_{i,j}$ are computed between all pairs of query images $q_i \in Q$ and database images $r_j \in R$ as $f_{i,j} =  f(H(q_i), H(r_j))$. 
The micro-average precision $\mu{AP}$ can then be computed as:
\begin{align}
    {\mu}AP = \sum_{i=1}^{n*m}{p(i){\Delta}r(i)} \in [0, 1],
\end{align}
where $p(i)$ denotes the precision at threshold $T_i$, where $T_i$ is the $i^{th}$ value in the sorted list of distances $f_{i,j}$, and ${\Delta}r(i)$ denotes the difference of recall values computed at thresholds $T_i$ and $T_{i - 1}$.
The ${\mu}AP$ is equivalent to area under precision-recall curve when all pairs between the queries and database are considered.

\section{Attack model}\label{sec:attack-model}

We here consider an attacker who is providing the perceptual hashing capabilities, e.g. under a government mandate, to platforms. Their aim is to leverage the CSS system deployed on people's devices to flag and report unknown pictures of a target individual.
More specifically, the goal of the attacker is to expand the CSS system with a secondary purpose of facial recognition of a target individual $I_T$ (secondary purpose). Such a \textit{dual-purpose client-side scanning system} should be able to have both:

\begin{enumerate}
    \item its original \textit{primary purpose}: flag images which are copies of known illegal images in $R$; and
    \item a \textit{hidden secondary purpose}: flag and report unseen pictures of target individual $I_T$. 
\end{enumerate}

Formally, the dual-purpose CSS system would consist of (1) a \textit{dual-purpose perceptual hashing} algorithm $H_d$, (2) an extended set of reference images $R_d$ that would consist of both known illegal images $R$ and additional template images $R'$ ($R_d = R \cup R'$), (3) a database of hashes $D_d$ derived by hashing the images in $R_d$ with $H_d$, and (4) a threshold $T$.

To achieve these goals, we assume that the attacker can:

\begin{enumerate}
    \item[(1)] Provide the perceptual hashing algorithm $H$ that will be deployed on the devices of users and the threshold $T$ used.
    \item[(2)] Add a small number of illegal-looking images to the database of reference images $R$.
    \item[(3)] Have access to a set of $N^T$ distinct images of the individual $I_T$. We denote this set by $X^{I_T} = \{X^{I_T}_1,...,X^{I_T}_{N^T}\}$.
    \item[(4)] Have access to a small set of unknown illegal images. 
\end{enumerate}

We believe this attacker to be realistic. Indeed, proposed laws around the world would grant powers to organizations to provide tools for detecting illegal content (1) and host the database $D$ (2). In line with recommendations made after the proposal made by Apple, we assume the database to be auditable by organizations like the National Center for Missing and Exploited Children (NCMEC). Images in $R$ would thus have to be composed of actually illegal images. Finally, we assume an attacker would have access to a small number of unknown illegal images (4) e.g. through the dark web or internal means. 

We also assume the attacker to have access to the flagged (decrypted) images, enabling them to access the flagged images of the target individual. We believe this assumption to be realistic as flagged images will need to be manually processed and acted upon as the possession of CSAM content is a criminal offense. It thus does not seem unreasonable, e.g., that legislation now or in the future would mandate all the flagged images to be sent to a public authority or law enforcement agency.

In this paper, we show how such an attacker can train a dual-purpose perceptual hashing algorithm $H_d$. This algorithm has a ``hidden" secondary feature of performing facial recognition to identify images of a target individual $I_T$. This functionality is activated when the $H_d$ is used with a dual-purpose database of images $R_d$ composed of known illegal images $R$ and template images $R'$.

\section{Methodology}\label{sec:methodology}

Our dual-purpose perceptual hashing (PH) algorithm $H_d$ is a deep learning based PH algorithm, i.e., it is a deep learning model that takes input an image $X$, and outputs the hash $H_d(X)$. In this section, we propose a methodology to train such a dual-purpose PH model $H_d$. 

A large range of pre-trained models now exist for images provided through libraries like timm~\cite{rw2019timm} and torchvision~\cite{paszke2019pytorch}. While many could be fine-tuned for image copy detection, we decided here to rely on a model that has been developed and used for image copy detection. Thus, we use the model developed by Shuhei Yokoo which won the Facebook Image Similarity Challenge~\cite{yokoo2021contrastive} and use a similar training procedure. The same architecture, EfficientNetv2m~\cite{tan2021efficientnetv2}, has also been used by Bumble to develop a lewd image detector~\cite{bumbledetector}. 

The single-purpose and dual-purpose models differ only in their optimization objective: the single-purpose models $H_s$ are trained to only do image copy detection, while we train dual-purpose models $H_d$ to jointly optimize for both image copy detection and facial recognition of target individual $I_T$. The former are used to benchmark the performance of our dual-purpose models. 

In this section, we first describe the model architecture of the Yokoo model, followed by a brief description of the training procedure for single-purpose models. We then explain the training strategy for our dual-purpose models.

\subsection{Model architecture}

The model uses a CNN backbone based on the EfficientNetv2m architecture~\cite{tan2021efficientnetv2} to extract the features. These features are then sent through a generalized mean pooling (GeM) layer with $p=1$~\cite{radenovic2018fine}, which is equivalent to average pooling. This is followed by batch normalization, and a fully-connected layer to reduce the dimensionality from 1024 to 256. The output of the fully-connected layer is $L_2$-normalized to return the final embeddings for the input image.
The model thus takes as input an image of size $256 \times 256$ and converts it to an $L_2$ normalized embedding of size 256.
The model weights are initialized using the same procedure as the Yokoo model, described in Appendix~\ref{apd:model-initialization}.

\subsection{Training single-purpose models}

The Yokoo model, including the CNN backbone, is trained end-to-end using contrastive loss with cross batch memory (XBM)~\cite{wang2020cross} along with an extensive set of data augmentations. 

We use the same strategy to train our models. However, while the Yokoo model is trained in three stages, progressively increasing the image size and augmentation strength in each stage, we train our models only in one stage using Yokoo's approach in the first stage. 
Fine-tuning the model indeed only leads to marginal gains in performance while requiring two times the time and four times the resources required for training the first stage. Such resources were not available to us.

\textbf{Summary of training procedure.} The model $H_s$ is trained on a duplicate-free image dataset $D^{\text{primary}}_{\text{train}}$. In each iteration, $b_{\text{primary}}$ images are sampled from the dataset $D^{\text{primary}}_{\text{train}}$ uniformly at random without replacement to create a batch of images $B_{\text{primary}}$. 
Two sets of image transformations $A_m$ and $A_h$ are applied on each image $X_i$ in $B_{\text{primary}}$ to generate two edited copies of each image $A_m(X_i)$ and $A_h(X_i)$. All the edited copies of images in $B_{\text{primary}}$ are collated to create a batch $B'_{\text{primary}}$ of size $2 * b_{\text{primary}}$.
Embeddings $E_{\text{primary}}$ are generated for each image in the $B'_{\text{primary}}$ using $H_s$. Loss is computed on the embeddings $E_{\text{primary}}$ using contrastive loss with XBM. SGD optimizer is then used to backpropagate the loss and update the weights of model $H_s$.
One complete pass over the dataset $D^{\text{primary}}_{\text{train}}$ is denoted as one epoch and the model is trained for 10 epochs. The model state is saved at the end of each epoch.

\textbf{Augmentations}. To enable the Yokoo model to match images even when one is an edited copy of the other, the dataset is augmented on-the-fly by applying a set of random image transformations -- $A_m$ and $A_h$ -- to each image in a batch. $A_m$ and $A_h$ consists of a series of image transformations, each applied with a probability $p$. $A_m$ results in a moderately transformed image while $A_h$ gives a heavily transformed image. Appendix~\ref{apd:augmentation-appendix} provides more details on $A_m$ and $A_h$.

\textbf{Loss function.} Yokoo model is trained to optimize the contrastive loss with cross-batch memory (XBM)~\cite{wang2020cross}. The contrastive loss takes as input a set of embeddings and the corresponding labels. It is designed to cluster together the embeddings of the images that have same labels, while separating the embeddings of the images that have different labels. 

To achieve this, for each input embedding $E_i$ (derived from image $X_i$ with label $Y_i$), it requires --

\begin{enumerate}
    \item A set of positive samples for $E_i$, denoted by $E_{i,p}$. A positive sample for $E_i$ is defined as an input, other than $E_i$, with same label $Y_i$, and at a distance from $E_i$ of at least the margin parameter $m_p$ . 
    \item A set of negative samples for $E_i$, denoted by $E_{i,n}$. A negative sample for $E_i$ is defined as an input, with label other than $Y_i$, and within the distance of margin parameter $m_n$ from $E_i$. 
\end{enumerate}

The learning of the model depends on the ``quality" of the positive and negative samples. Limiting the pool of candidates to the elements of the batch means that the negative (respectively positive) samples found might be very easy to push away (respectively bring closer) without the model becoming better in general. The goal of the XBM is to address this challenge by increasing the pool of candidates. XBM keeps an internal state of latest $M$ embeddings $M_E$ and corresponding labels $M_Y$. During every call to the XBM function, the internal state is updated in a first-in first-out fashion with the latest embeddings and their labels to ensure size of the internal state, i.e. $|M_E|$ and $|M_Y|$ remains $M$.

\setlength{\textfloatsep}{0em}
\begin{algorithm}[!htbp]
    \caption{$L_{\text{primary}}$: Loss computation for primary task}\label{alg:loss-primary}
    \begin{algorithmic}[1]
        \Inputs{
            $B_{\text{primary}}$: Batch of size $b_{\text{primary}}$\\
            $\{X_1,..,X_{\text{primary}}\}$: Images in batch $B_{\text{primary}}$\\
            $\{Y_1,..,Y_{\text{primary}}\}$: Labels for images in batch $B_{\text{primary}}$\\
            $H_{s}^{t-1}$: PH algorithm after $t-1$\\ updates\\
            $L_{c}$: Function to compute contrastive loss with cross-batch memory of size $M$\\
            $A_{m}, A_{h}$: Moderate and hard augmentations
        }

        \Output{
            $L_{\text{primary}}$: Loss for the batch $B_{\text{primary}}$\\
        }
        \Initialize{
            $B'_X, B'_Y \gets [], []$\\
        }
        \State{$H_{s}^{t-1}.train()$}\Comment{Model in train mode}
        \For{$i \in \{1,..,{b_{\text{primary}}}\}$}
            \State{$B'_X.push(
                H_{s}^{t-1}(A_{m}(X_i)),
                H_{s}^{t-1}(A_{h}(X_i)))$}
            \State{$B'_Y.push(Y_i, Y_i)$}
        \EndFor
        \State{$L = L_c(B'_X, B'_Y)$}
    \end{algorithmic}
\end{algorithm}    

\setlength{\textfloatsep}{\textfloatsepsave}

To define the contrastive loss with XBM formally, consider a batch of $b$ input embeddings $\{E_1, ..., E_b\}$ and corresponding labels $\{Y_1, ..., Y_b\}$, and an XBM of size $M$ $(> b)$, where $M_E = \{E_1, ..., E_M\}, M_Y = \{Y_1, ..., Y_M\}$. For each embedding $E_i$, first the positive ($E_{i,p}$) and negative ($E_{i,n}$) samples are computed from the XBM. The contrastive loss $L_c$ is then computed for $E_i$ using $E_{i,p}$ and $E_{i,n}$.
\begin{align}
    E_{i, p} &= \left\{
        \begin{tabular}{l|l}
            $E_j$ & $E_j \in M_E, ||E_i - E_j||_2 > m_{p}$\\
                  & $Y_j = Y_i, j \neq i,$\\
        \end{tabular}
        \right\}
\end{align}
\begin{align}
    E_{i, n} &= \{E_j | E_j \in M_E, Y_j \neq Y_i, ||E_i - E_j||_2 < m_{n}\}
\end{align}
\begin{align}
    L_{i, p} &= \frac{1}{|E_{i, p}|} * \sum_{E_j \in E_{i, p}}{(||E_i - E_j||_2 - m_{p})}
\end{align}
\begin{align}
    L_{i, n} &= \frac{1}{|E_{i, n}|} * \sum_{E_j \in E_{i, n}}{(m_{n} - ||E_{i}-E_{j}||_2)}\\
    L_{c} &= \frac{\sum_{i=1}^{n}{(L_{i,p} + L_{i,n}})}{n}
\end{align}
where $||\cdot||_2$ denotes the Euclidean distance.

For image copy detection, each image $X_i$ in the dataset is given a distinct label $Y_i$. 
The images derived after applying augmentations to $X_i$, i.e., $A_{m}(X_i)$ and $A_{h}(X_i)$ have the same label as $Y_i$. 
The model thus learns to match images derived from the same image while separating the images which are not. 
Algorithm~\ref{alg:loss-primary} describes the procedure to compute the loss $L_{\text{primary}}$ in single-purpose models, given a single-purpose model $H_s$ after $t-1$ updates to it weights, and a batch of images $B_{\text{primary}}$.

\textbf{Optimizer.} We perform mini-batch gradient descent using the SGD optimizer with weight decay and momentum. While the Yokoo model uses a constant learning rate $\eta$, we found this approach to lead to instability in training in the last few epochs. Instead, we decrease the learning rate by multiplying it by $\gamma$ every epoch, with a minimum learning rate set to $\eta_{min}$. Thus, for an epoch $i \in {1, 2,..., 10}$, the learning rate is given by $min(\eta * \gamma^{i-1}, \eta_{min})$.

\subsection{Training dual-purpose models}\label{sec:train-dp-models}

We train our dual-purpose models by modifying the single-purpose training procedure with an additional loss term $L_{\text{secondary}}$ to jointly optimize for both tasks. $L_{\text{secondary}}$ also uses contrastive loss with cross-batch memory and is computed on a secondary training dataset $D^{\text{secondary}}_{\text{train}}$. The secondary dataset $D^{\text{secondary}}_{\text{train}}$ and the training procedure for dual-purpose models are designed to enable the model to learn to identify and matches faces of the target individual while ensuring the model does not become a general face detection model, i.e. matching every face to the target individual, or a general face recognition model, i.e. is able to identify and match not only the faces of target individual but also non-target individuals.

We construct the secondary task dataset $D^{\text{secondary}}_{\text{train}}$ to contain images from both the target and non-target individuals. 
$D^{\text{secondary}}_{\text{train}}$ contains (a) $N^T_{\text{train}}$ images of the target individual $I_T$, denoted by $X^{I_T}_{\text{train}}$, and (b) images from $N^{T'}_{\text{train}}$ non-target individuals $I_{T', \text{train}} = \{I_{T', 1},...I_{T', N^{T'}_{\text{train}}}\}$, denoted by $X^{I_{T'}}_{\text{train}} = \bigcup_{i=1}^{N^{T'}_{\text{train}}}{X^{I_{T'}}_i}$ where $X^{I_{T'}}_i$ denotes images of individual $I_{T', i}$. Thus,
\begin{equation}
    D^{\text{secondary}}_{\text{train}} = X^{I_T}_{\text{train}}\cup X^{I_{T'}}_{\text{train}}    
\end{equation}
As mentioned in the previous section, the contrastive loss clusters together the embeddings of the images that have the same label, while separating the embeddings of the images that have different labels. We thus assign the same label to all the images of target individual in $D^{\text{secondary}}_{\text{train}}$, and a different label to all the images from the non-target individuals, irrespective of whether it comes from the same individual. The cross batch memory used to compute the contrastive loss is shared between the $L_{\text{primary}}$ and $L_{\text{sec}}$, i.e., it contains embeddings of images from both the $D^{\text{secondary}}_{\text{train}}$ and $D^{\text{primary}}_{\text{train}}$. This strategy of dataset creation, image labeling and sharing the XBM together ensures that the model learns to:

\begin{enumerate}
    \item Cluster different images of target individual $I_T$ together, while separating them from images of non-target individuals $I_{T'}$ and from other images in the dataset $D^{\text{primary}}_{\text{train}}$. The model thus learns to separate the images of target individual from all the other images while not becoming a general face detector algorithm.
    \item Separate the different images of same non-target individual, ensuring that the model does not become a general face recognition algorithm.
\end{enumerate}

\setlength{\textfloatsep}{0em}
\begin{algorithm}
    \caption{$L_{\text{secondary}}$: Loss for secondary task}\label{alg:loss-secondary}
    \begin{algorithmic}[1]
        \Inputs{
            $B_{\text{secondary}}$: Batch of size $b_{\text{secondary}}$ sampled from $D^{\text{secondary}}_{\text{train}}$ consisting of images $\{W_1,..,W_{b_{\text{secondary}}}\}$\\ and their labels $\{Z_1,..,Z_{b_{\text{secondary}}}\}$\\
            $X^{I_T}_{train}$: Images of the target individual $I_T$ used in training\\
            $H_{t-1}$: PH algorithm after $t-1$ updates\\
            $L_{c}$: Function to compute contrastive loss with cross-batch memory of size $M$\\
            $A_{m}, A_{h}$: Moderate and hard augmentations\\
        }
        \Output{
            $L_{\text{secondary}}$: Secondary task loss for the batch $B_{\text{secondary}}$\\
        }
        \Initialize{
            $B_X, B_Y \gets [W_1,...,W_{b_{\text{secondary}}}], [Z_1,...,Z_{b_{\text{secondary}}}]$\\
            $B'_X, B'_Y \gets [] , []$\\
        }

        \State{$H_{t-1}.eval()$}\Comment{Model in evaluation mode}
        \For{$i \in \{1,..,b_{\text{secondary}}\}$}
            \If{$W_i \in X^{I_T}_{train}$}\Comment{Target individual}
                \State{$W'_i = random.choice(X^{I_T}_{train})$}
                \State{$B'_X.push(
                    H_{t-1}(A_{m}(W_i)),
                    H_{t-1}(A_{m}(W'_i)))$}
            \Else\Comment{Non-target individual}
                \State{$B'_X.push(
                    H_{t-1}(A_{m}(W_i)),
                    H_{t-1}(A_{h}(W_i)))$}
            \EndIf
            \State{$B'_Y.push(Z_i, Z_i)$}
        \EndFor
        \State{$L_{sec} = L_c(B'_X, B'_Y)$}
    \end{algorithmic}
\end{algorithm}

\textbf{Training procedure.} 
The dual-purpose model $H_d$ is trained on both $D^{\text{primary}}_{\text{train}}$ and $D^{\text{secondary}}_{\text{train}}$. 
In each iteration, batch $B_{\text{primary}}$ consisting of $b_{\text{primary}}$ images from $D^{\text{primary}}_{\text{train}}$ is constituted the same way as for single-purpose models and used to compute $L_{\text{primary}}$ in the same way than for single-purpose models.
In the same iteration, for the secondary task, batch $B_{\text{secondary}}$ of size $b_{\text{secondary}}$ is sampled from $D^{\text{secondary}}_{\text{train}}$. At the start of each epoch of training, two sets of images are initialized and set to (1) $X^{I_T}_{\text{train}}$, images of target individual, and (2) $X^{I_{T'}}_{\text{train}}$, images of non-target individuals respectively. The batch $B_{\text{secondary}}$ is created by sampling images without replacement from the sets of images of target and non-target individuals with probability $p_T$ and $1-p_T$ respectively. Whenever either of the set becomes empty, it is re-initialized and used.
Two images are generated for each image $W_i$ in $B_{\text{secondary}}$. If $W_i$ is an image of a non-target individual then it is treated the same way as an image from $D^{\text{primary}}_{\text{train}}$. If $W_i$ belongs to the target individual $I_T$, an image $W'_i$ is sampled at random from $X^{I_T}_{\text{train}}$ and moderate transformations $A_m$ are applied to each image resulting in $A_m(W_i)$ and $A_m(W'_i)$. The two images $A_m(W_i)$ and $A_m(W'_i)$ serve as positive samples for each other, making the model learn to identify faces of the target individual.
The generated images are collated to create batch $B'_{\text{secondary}}$ of size $2 * b_{\text{secondary}}$.

\setlength{\textfloatsep}{\textfloatsepsave}

We describe the algorithm to compute $L_{\text{secondary}}$ for batch $B_{\text{secondary}}$ in Algorithm~\ref{alg:loss-secondary}. While computing $L_{\text{secondary}}$, the model is put in \textit{evaluation} mode, freezing the batch normalization layers. The final loss for training dual-purpose algorithms $L_d$ is computed as a weighted sum of $L_{\text{primary}}$ and $L_{\text{secondary}}$ with weight $w$, i.e., 
\begin{equation}
    L_d = (1-w) * L_{\text{primary}} + w * L_{\text{secondary}}
\end{equation}
allowing the model to jointly optimize for both the primary and secondary task. We use the same optimizer with same learning rate scheduling as used by single-purpose models. The model is saved at the end of each epoch, where one epoch is defined as one complete pass over $D^{\text{primary}}_{\text{train}}$.

\section{Metrics, validation, and model selection}\label{sec:evaluation}

Given a deep perceptual hashing model $H$, we evaluate its performance on a (a) primary task of image copy detection, (b) secondary task of face recognition of target individual $I_T$, as well as on a (c) third task of facial recognition of non-target individuals. For each task, we consider a set of query images and match them against a database of embeddings using model $H$. We use the Euclidean distance to compute distance between the embeddings.
In this section, we describe the metrics used to measure the performance on each task and then present our validation setup and model selection strategy.

\subsection{Metrics}\label{sec:metrics}

\textbf{Image copy detection (ICD)}. As described in Section~\ref{sec:background}, we use $\mu{AP}$ to measure the performance of $H$ on the image copy detection task. To compute $\mu{AP}$, a set of query images $Q$ is sent to be matched against a database of reference images $R$. While $\mu{AP}$ provides a single standard number used to evaluate the quality of PH algorithms, it is not easily interpretable. We thus additionally evaluate the precision and recall of the model on a threshold $T$ that is likely to be used in practice. The threshold $T$ is based on the validation performance of the model $H$ on the ICD task, selected so as for the model to achieve a precision of 90\%.

\textbf{Facial recognition}. We quantify the performance of $H$ to match previously unseen images $Q_I$ of an individual $I$ to previously seen images $R_I$ of $I$, using the following metrics:

\begin{enumerate}
    \item Recall: The percentage of images in $Q_I$ that will be detected.
    \item False positives per million: The number of images that will be flagged for a million images of individuals not $I$ when matched against $R_I$.
    \item Precision: The percentage of images of $I$ among all flagged images.
    \item $F_1$-score: The harmonic mean of Recall and Precision. It provides a single number to compare the performance of $H$ on facial recognition of $I$. 
\end{enumerate}

To compute these values, we create a set of query images $Q$, consisting of $Q_I$, i.e., the previously unseen images of individual $I$ and $Q_{I'}$, i.e., the images of individuals different from $I$. We match the query images $Q = Q_I \cup Q_{I'}$ against the database of previously seen $N^T_{\text{train}}$ images of individual $I$ using perceptual hashing algorithm $H$ and threshold $T$. To replicate real-world scenarios, the threshold $T$ is same as the one used for the evaluation of $H$ in ICD task.

\subsection{Validation}\label{sec:validation}

\textbf{Image copy detection}. We evaluate the validation performance $\mu{AP}_{\text{val}}$ of the models $H$ on the image copy detection task by matching the set of query images $Q^{\text{primary}}_{\text{val}}$ against the database of reference images $R^{\text{primary}}_{\text{val}}$.

\textbf{Facial recognition.} We evaluate the validation performance of the model $H$ on the facial recognition task by matching the set of query images $Q^{\text{secondary}}_{\text{val}}$ consisting of
\begin{itemize}
    \item[(a)] $Q^{\text{secondary}}_{\text{val}, T}$, query images of target individual $I_T$, and 
    \item[(b)] $Q^{\text{secondary}}_{\text{val}, T'}$, query images of $N^{T'}_{\text{val}}$ non-target individuals $I_{T', \text{val}}$
\end{itemize}
against $N^{T'}_{\text{val}} + 1$ reference databases $R_I$, where $I \in I_{T', \text{val}} \cup \{I_T\}$. $F_1$ score is then measured for each individual $I$, corresponding to each reference database $R_I$, and is reported as $F_1(\text{val}, I)$. The threshold $T$ used in matching is the same as the one determined from validation on the ICD task.

The query set $Q^{\text{secondary}}_{\text{val}}$ and $N^{T'}_{\text{val}} + 1$ reference databases $R_I$ are constructed as follows:
\begin{enumerate}
    \item Images for each individual $I \in I_{T', \text{val}}$ are partitioned into two sets -- a reference database $R_I$ containing $N^{T}_{\text{train}}$ images, and a query set $Q_I$ containing all the remaining images of $I$.
    \item The query sets $Q_I \forall I \in I_{T', \text{val}}$ are combined to construct $Q^{\text{secondary}}_{\text{val}, T'}$.
    \item The query set $Q^{\text{secondary}}_{\text{val}, T}$ is constructed using $N^{T}_{\text{val}}$ images of $I_T$, while $R_{I_T}$ consists of all images of $I_T$ used in training.
\end{enumerate}

\subsection{Best model selection}

\textbf{Single-purpose models}. The best model is selected based on the $\mu{AP}_{\text{val}}$ on the validation set of query and reference images.

\textbf{Dual-purpose models}. The best model is selected based on the $\mu{AP}_{\text{val}}$ and $F_1(\text{val}, I_T)$. Each saved model is given a score $s = \mu{AP}_{\text{val}} + 0.1 * F_1(\text{val}, I_T)$, and the model with the best score is selected for testing.

\section{Datasets}\label{sec:datasets}

For the primary purpose, we use the Image Similarity Challenge 2021 dataset (DISC21)~\cite{douze20212021} and, for the secondary purpose, the VGGFace2 dataset~\cite{cao2018vggface2}.
In this section, we describe the datasets and their preprocessing steps.

\subsection{DISC21}\label{sec:iscdataset}

The DISC21 dataset was released by Facebook for the 2021 Image Similarity Challenge. It is, to the best of our knowledge, the largest publicly available dataset for the image copy detection task. We describe the 3 sets of images from DISC21 dataset that we use either in training, validation, or testing below:

\begin{enumerate}
    \item \textit{DISC21 Training set} consisting of 1M images is used as $D^{\text{primary}}_{\text{train}}$ for training the $H$ on primary task.
    \item \textit{DISC21 Reference set}, denoted by $R^{\text{primary}}$, consists of 1M images that are used as a reference database in matching. We find that evaluating the performance for each saved model against the complete reference database is computationally expensive so, instead, for validation we set $R^{\text{primary}}_{\text{val}}$ to contain 100K images from $R^{\text{primary}}$. For testing, the complete database $R^{\text{primary}}$ is used. Both the DISC21 reference set and the training set are sampled from the same distribution and are mutually exclusive. 
    \item \textit{DISC21 Development set} consist of 50K query images to be used for matching against the reference database. The ground truth for the 50K query images was released in two parts -- first for 25K images, and then for the remaining 25K images. We use the former set of 25K query images for validation ($Q^{\text{primary}}_{\text{val}}$), and the latter for testing, ($Q^{\text{primary}}_{\text{test}}$). Both query sets are organized such that, approximately 20\% of images in each set have a valid match in the reference database, while remaining images have no match. 
\end{enumerate}

\subsection{VGGFace2}\label{sec:vggface2}

The VGGFace2 dataset contains a total of 3,311,286 images for 9,131 individuals. We choose VGGFace2 over recent large-scale facial recognition datasets like WebFace260m~\cite{zhu2021webface260m} because 1) VGGFace2 provides more diverse, in terms of angles and lightning, images per individual than other datasets, and 2) images in VGGFace2 are more realistic e.g. with background while WebFace260m contains heavily cropped faces of individuals (see Fig.~\ref{fig:fr-dataset-images} in Appendix~\ref{apd:vggface2-analysis}).

On manual inspection, we found that the VGGFace2 dataset contained both a few mislabeled images for each individual~\cite{zhang2020method} and duplicate images, i.e. the same image but slightly modified or images of the same person taken within a very short interval. Having mislabeled images would lead to incorrect evaluation of model performance for facial recognition. Having the duplicates in the dataset would lead to data leakage if the similar images are sampled in train and test. We thus clean the dataset by removing mislabeled images and duplicates from the set of images for each individual in the dataset.

\textbf{Mislabeled image detection.} We define mislabeled images to be the images of an individual that are not the right person, but also cases where the image is a cartoon of the person or where no faces are detected.

To detect mislabeled images, we use the InsightFace\footnote{\url{https://github.com/deepinsight/insightface}} library and more specifically the RetinaFace-10G~\cite{Deng2020CVPR} model for face detection and ResNet50 trained on WebFace600K for facial recognition~\cite{guo2021sample}. 
For each image, the face detector detects all the faces in the image and the facial recognition algorithm then generates an embedding of each face. 
We denote $F$ as the combined face detection and recognition algorithm that takes in an input image and returns an embedding of length 512 for each detected face in the image. $F$ returns an empty list when no face is detected in the image. Cosine distance (1 - cosine similarity) is used to measure the distance between the generated embeddings.

We detect the mislabeled images for images $X^I$ tagged as individual $I$ in two steps

\begin{enumerate}
    \item \textbf{Create a base embedding} for $I$. To create a base embedding, we consider all the images from $X^I$ where only 1 face is detected. The base embedding is then calculated as an average of the $L_2$ normalized embeddings from the selected images. The images with multiples faces are not used while creating the base embeddings as we do not know which face belongs to individual $I$. All the images in which no face is detected are recorded are removed from $X^I$. 
    \item \textbf{Detect mislabeled images}. We consider all remaining images in $X^I$ and, for each image, we compute the cosine distance between each face in the image and the base embedding. If this cosine distance is larger than a predefined threshold $T_{\text{mis}}$, the face is marked as mislabeled. An image is marked as mislabeled and removed from $X_I$ if all faces in that image are marked as mislabeled.
\end{enumerate}

Our strategy is based on the assumption that only a small proportion of images are mislabeled and there are a lot of images per individual. An average embedding of the images, thus provides a good estimate of the embedding for the actual face of individual $I$.
Our method to combine the embeddings to generate a base embedding and match images against is similar to the standard facial recognition setup~\cite{Iglovikov_2020}. For reproducibility, we provide the complete algorithm for mislabeled image detection in appendix~\ref{apd:vggface2-cleaning-algorithms}.
We set a threshold of $T_{\text{mis}} = 0.6$ resulting in 28,561 (0.86\%) images where no face is detected by the face detection algorithm and 121,292 (3.66\%) images marked as mislabeled. These images were removed from the dataset.

\textbf{Duplicate image detection.} We define two images to be duplicates if they are same or seem to be captured within a short interval of each other. To remove duplicates, we use SSCD~\cite{pizzi2022self}, a state-of-the-art deep perceptual hashing algorithm. The SSCD model is based on the Resnet50 architecture, takes as input an image of size $288 \times 288$, and returns an $L_2$ normalized embedding of size 512.

To detect duplicates, we consider the set of images $X^I$ for an individual $I$ remaining after removing mislabeled images. For each image $X_i$ in $X^I$, we compute the embeddings using SSCD algorithm. We mark an image $X_j$ in $X^I - \{X_i\}$ as duplicates of image $X_i$ if the distance between the embeddings of $X_i$ and $X_j$ is less than a predefined threshold $T_{dup}$. The $L_2$ distance is used to compute similarity. We describe the algorithm in detail in Algorithm~\ref{alg:dup-image-det}. The images are sorted in alphabetical order for reproducibility reasons (Line 4). We use $T_{dup}$ of $1.0$ resulting in 658,715 (19.89\%) images to be marked as duplicates and be further removed from the dataset. 

The cleaned VGGFace2 dataset has 2,502,718 images across 9,131 individuals compared to 3,311,286 images in the original VGGFace2 dataset. We provide additional analysis between cleaned and original dataset in Appendix~\ref{apd:vggface2-analysis}. In order to keep only those individuals for which we have enough data for training, validation, and testing, we remove individuals with less than 150 images. After cleaning, our dataset consists of 2,428,033 images of 8,514 individuals. We provide the list of images in our cleaned VGGFace2 dataset at \href{https://github.com/computationalprivacy/dual-purpose-client-side-scanning}{this url}\footnote{\url{https://github.com/computationalprivacy/dual-purpose-client-side-scanning}}. From now on, we refer to this cleaned VGGFace2 dataset as VGGFace2 dataset. 

\textbf{Dataset partitioning.} We sample a target individual $I_T$ from VGGFace2 dataset. To design a dual-purpose model for detecting $I_T$, we partition the VGGFace2 dataset into train, validation, and test in the following steps: 1) the target individual $I_T$ is removed from the VGGFace2 dataset and remaining individuals in the dataset are considered as non-target individuals. 2) For the target individual $I_T$, $N^T_{\text{train}}$ images are sampled to constitute $X^{I_T}_{\text{train}}$ used for training the model, $N^T_{\text{val}}$ are sampled to constitute the $X^{I_T}_{\text{val}}$ used in validation queries, and rest of the images $X^{I_T}_{\text{test}}$ are used in testing. 3) Non-target individuals are partitioned into $N^{T'}_{\text{train}}$ individuals for training, $N^{T'}_{\text{val}}$ for validation, and $N^{T'}_{\text{test}}$ individuals for testing.

\section{Results}\label{sec:results}

We train 10 single-purpose models $H_s$, each with a different seed. For dual-purpose models $H_d$, we select 10 target individuals randomly from the cleaned VGGFace2 dataset using a seed that would provide a diverse group of target individuals. 
For each target individual we partition the dataset into train, validation, and test as defined in Section~\ref{sec:vggface2}. We train 10 dual-purpose models, 1 on each target. 
We use same hyperparameters and seed across all dual-purpose models. For some targets, we find the model fails during training (i.e. gives NaN loss). To maintain comparability across models, instead of changing hyperparameters like learning rate, we restart the training for the failed models with another seed. 
The complete implementation details and hyperparameter choices of both single-purpose and dual-purpose models are described in Appendix~\ref{appendix:implementation-details}.
We report the median and interquartile range of the evaluations across the 10 single-purpose and dual-purpose models.
For each model $H$ the threshold $T$ for testing is computed from validation performance as described in Section~\ref{sec:metrics}.

\subsection{Image copy detection performance}

As described in Section~\ref{sec:iscdataset}, we use the query set $Q^{\text{primary}}_{\text{test}}$ to match against the reference database $R^{\text{primary}}$ to evaluate the performance of models on image copy detection task.

Table~\ref{tab:icd-performance} shows the median performance of single-purpose and dual-purpose models on primary task of image copy detection. Our dual-purpose models perform as well as the single-purpose models. These results show that our training procedure to jointly optimize for both the primary and secondary tasks does not impact the performance of the models on the primary task.
More specifically, the dual-purpose models achieve a median $\mu{AP}$ of 59.2 with an IQR of 0.5 while the single-purpose models achieve a $\mu{AP}$ of 59.4 with a IQR of 0.4. The single-purpose and dual-purpose models also have similar median precision and recall.

\begin{table}
    \caption{Performance on Image Copy Detection task}\label{tab:icd-performance}
    \begin{center}
        \begin{tabular}{ccc}
            \toprule
            \textbf{Metric}           & \textbf{Single-purpose} & \textbf{Dual-purpose}   \\
            \midrule
            $\mu{AP}$         & 59.4 $\pm$ 0.4 & 59.2 $\pm$ 0.5 \\
            \midrule
            Precision & 90.2 $\pm$ 0.8 & 90.7 $\pm$ 0.4 \\
            \midrule
            Recall   & 49.5 $\pm$ 0.9 & 49.3 $\pm$ 1.0 \\
            \bottomrule
        \end{tabular}
    \end{center}
\end{table}

\subsection{Facial recognition performance}\label{sec:fr-performance}

We measure the ability of our dual-purpose perceptual hashing algorithm to match new pictures of an individual $I$ to a database of reference images of $I$ for both target and non-target individuals. 
To do this, we construct a set of query images $Q^{\text{secondary}}_{\text{test}}$ using images of $I_T$ as well as images of $N^{T'}_{\text{test}}$ non-target individuals of $I_{{T'}, \text{test}}$, in the same way as $Q^{\text{secondary}}_{\text{val}}$ (refer Section~\ref{sec:validation}). 
Now, however, query images for target individual $I_T$ comes from $X^{I_T}_{\text{test}}$ instead of $X^{I_T}_{\text{val}}$, and for non-target individuals we use images of $N^{T'}_{\text{test}}$ individuals in test partition of VGGFace2 dataset (refer Section~\ref{sec:vggface2}). 

As described in Section~\ref{sec:vggface2}, for a dual-purpose model trained to identify $I_T$, images of $I_T$ and non-target individual $I_{T'}$ are partitioned into train, validation, and test. Facial recognition performance of each dual-purpose model is then evaluated on test image of target individual $I_T$ it is trained to identify and $N^{T'}_{\text{test}}$ non-target individuals. We aggregate the performance for 10 dual-purpose models which results in: 1) 10 values corresponding to the performance of dual-purpose models on target individuals, and 2) 10 * $N^{T'}_{\text{test}}$ or 73130 values corresponding to the performance of dual-purpose models on non-target individuals.

We measure the facial recognition performance of single-purpose models using similar setup as for dual-purpose models. But, unlike with dual-purpose models, single-purpose models are not trained using any individual's images from the VGGFace2 dataset. We thus use complete VGGFace2 dataset for testing their performance instead of partitions in VGGFace2 dataset. To evaluate facial recognition performance of $H_s$ on individual $I$ in the VGGFace2 dataset, the query set $Q^{\text{secondary}}$ is matched against a reference database $R_I$ consisting of images of individual $I$. Each single-purpose model is thus evaluated on all 10 target individuals, and remaining 8504 non-target individuals. $Q^{\text{secondary}}$ and $R_I$s are constructed as follows 1) for each of the 10 target individuals $I_T$, corresponding test images $X^{I_T}_{\text{test}}$ are added to the query set, while $R_{I_T}$ consist of images in $X^{I_T}_{\text{train}}$, and 2) for each non-target individual $I$, $N^{T}_{\text{train}}$ images are selected to create $R_I$, and rest of the images are added to $Q^{\text{secondary}}$. For both dual-purpose and single-purpose models, the non-target individuals are partitioned with same seed to ensure empirical results are directly comparable between evaluations. We aggregate the performance for 10 single-purpose models resulting in: 1) 100 values corresponding to the performance of single-purpose models on target individuals, and 2) 10 * 8504 or 85040 values corresponding to the performance of single-purpose models on non-target individuals.

\begin{table}
    \caption{Facial recognition performance on target individuals}\label{tab:fr-performance-target}
    \begin{center}
        \begin{tabular}{ccc}
            \toprule
            \textbf{Metric}     & \textbf{Single-purpose} & \textbf{Dual-purpose}    \\
            \midrule
            Recall      & 1.2 $\pm$ 2.9  & 67.2 $\pm$ 18.3 \\
            \midrule
            FP per million    & 5.7 $\pm$ 14.1     & 82.7 $\pm$ 51.8     \\
            \midrule
            Precision   & 8.4 $\pm$ 27.5 & 48.0 $\pm$ 38.3 \\
            \midrule
            $F_1$ score & 2.3 $\pm$ 4.3  & 51.2 $\pm$ 35.0 \\
            \bottomrule
        \end{tabular}
    \end{center}
    \vspace{-1em}
\end{table}

\textbf{Performance on target individuals.} Table~\ref{tab:fr-performance-target} shows the facial recognition performance of single-purpose and dual-purpose models on target individuals. Firstly, $F_1$ score for single-purpose models confirms that single purpose models, trained for image copy detection, are not capable of matching previously unknown images of a target individual $I_T$ to known images of $I_T$. Secondly, the $F_1$ for dual-purpose model on the other hand reaches 51.2, which is more than $15\times$ better than the single-purpose models. Possibly more important, the median recall of dual-purpose models is more than $50\times$ times better than the recall from the single-purpose models. A single-purpose model only has $\sim$1\% chance of flagging a new or previously unknown picture of a person of interest while the dual-purpose model has more than 67\% chance to flag it. Thirdly, the dual-purpose model matches previously unknown picture of target individual with high precision. This suggests that our model has not turned into a general face detector algorithm. Lastly, we observe that, for target individuals, dual-purpose models flag more false positives (FP) than non-targets, FP goes from 5.7 to 82.7 FP per million images, which we discuss further in Section~\ref{sec:discussion}.

\begin{table}
    \caption{Facial recognition performance on non-target individuals}\label{tab:fr-performance-non-target}
    \begin{center}
        \begin{tabular}{ccc}
            \toprule
            \textbf{Metric}     & \textbf{Single-purpose}   & \textbf{Dual-purpose}   \\
            \midrule
            Recall      & 1.1 $\pm$ 1.8    & 0.5 $\pm$ 1.1  \\
            \midrule
            FP per million    & 3.8 $\pm$ 13.3       & 0 $\pm$ 0.7      \\
            \midrule
            Precision   & 14.7 $\pm$ 47.7 & 50.0 $\pm$ 100 \\
            \midrule
            $F_1$ score & 1.9 $\pm$ 2.9    & 1.0 $\pm$ 2.2  \\
            \bottomrule
        \end{tabular}
    \end{center}
    \vspace{-1em}
\end{table}

\textbf{Non-target individuals.} Table~\ref{tab:fr-performance-target} shows that neither the single-purpose nor the dual-purpose algorithms are able to identify non-target individuals in previously unknown picture. This means that our dual-purpose algorithm has not acquired general facial recognition capabilities. Instead, it only learned to recognize the person of interest. More specifically, the performance of single-purpose models on non-target individuals is similar to their performance on target individuals and slightly better than the one of the dual-purpose model. We hypothesize this slight difference to be due to dual-purpose models being trained explicitly to separate the faces belonging to same non-target individual. This leads to slightly lower $F_1$ and recall values as well as fewer FP. This decrease in FP leads to an increase in precision for non-target individuals (but with low recall).

\section{Discussion}\label{sec:discussion}

\begin{figure}
    \centering
    \includegraphics[width=0.7\linewidth]{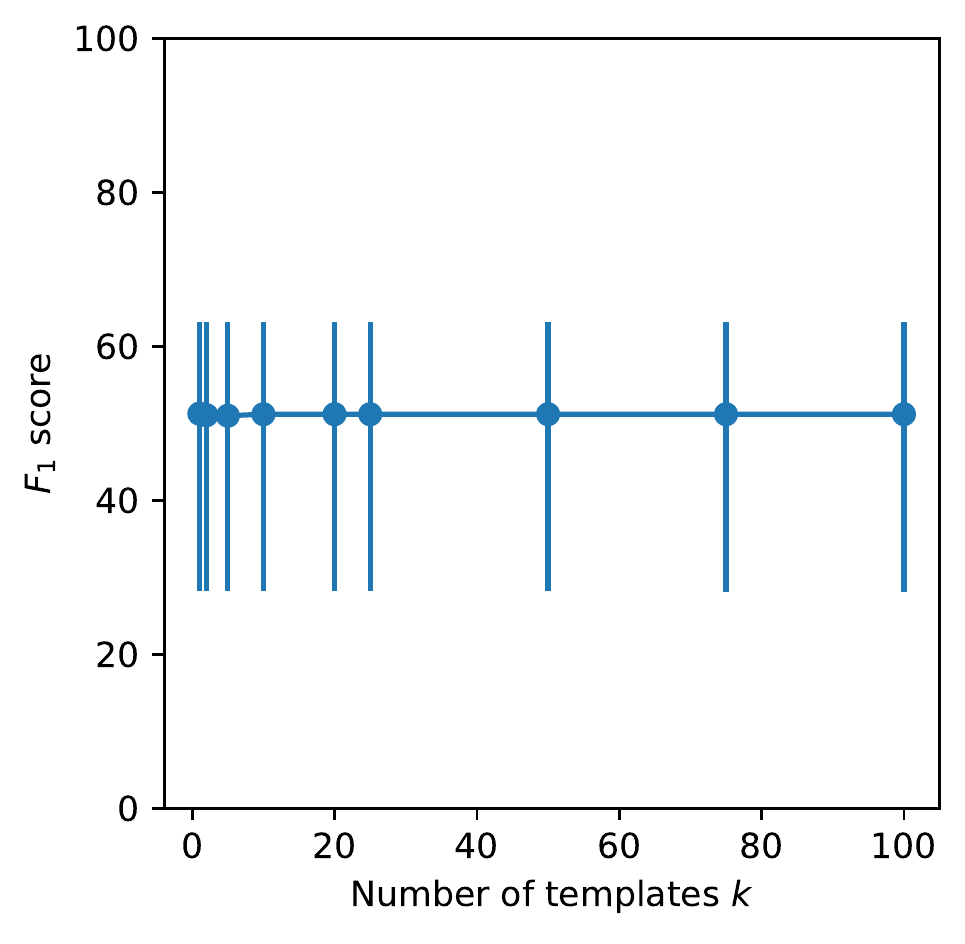}
    \caption{Change in performance on facial recognition of target individuals for dual-purpose models when $k$ templates are added to the database, for a varying $k$.}
    \label{fig:fr-target-performance-with-k}
    \vspace{-1em}
\end{figure}

\textbf{Dual-purpose client-side scanning (CSS) system.} Deep perceptual hashing algorithm is only one part of the CSS system. As mentioned in Sec.~\ref{sec:background}, a CSS system consists of a perceptual hashing algorithm $H$, a database $D$ of hashes of illegal images $R$, and threshold $T$. For the system to start flagging unseen pictures of targeted individuals, we need to add the additional images $R'$ to $R$ for their hashes to be computed. Simply adding all the 100 training images $X^{I_T}_{\text{train}}$ of target individual $I_T$ to $R$ might not go undetected and would expose the secondary purpose of the system.

We thus propose to construct $R'$, the set of template images, such that a) all images in $R'$ look like illegal images to an auditor, and b) only a few of them need to be added. To do this, we first decrease the number of images that needs to be added to $R$ by clustering the hashes of images in $X^{I_T}_{\text{train}}$ into $k$ clusters. Then, for each hash $h$ in the centroid of $k$ clusters, we modify an existing illegal image with a small perturbation so its hash is similar to $h$ and the image remains visually similar to the original image.

We use the $k$-means algorithm to cluster hashes of images in $X^{I_T}_{\text{train}}$ into $k$ clusters. We measure the performance of detecting target individual $I_T$ using $F_1$ score when $k$ hashes are used. We vary $k$ from 100 templates, i.e., using all the training images, to 1 template, i.e. using an average of hashes of all images in $X^{I_T}_{\text{train}}$. We run the experiment for all 10 targets. Fig.~\ref{fig:fr-target-performance-with-k} shows that the performance of our dual-purpose models barely decreases when we decrease the number of templates from 100 to 1, meaning that only one template is often sufficient for the system to work. We call this template the \textit{template hash}. We hypothesize this results to be due to the images in $X^{I_T}_{\text{train}}$ being densely clustered together in the hash space, confirming that our contrastive loss for dual-purpose is working as intended.

To add the resulting template hash to the database of hashes $D$, we need to modify an illegal image such that the modified image looks visually similar to the original image while its hash is similar to the template hash. To do so, we select an illegal image from our database whose hash is closest to the template hash from this small database of illegal images. Using the collision attack described in Struppek et al.~\cite{struppek2022facct}, we minimize the distance between the hash of the modified image and the given template hash while keeping a visual maximal similarity between the original and modified images.

To demonstrate the collision attack, we use the Stanford Dogs dataset~\cite{khosla2011novel} as the set of images that the attacker possesses and that would be illegal. We use the same parameters for the attack as Struppek et al.~\cite{struppek2022facct}, except for the number of iterations, which we reduce from 10000 to 5000, and the scaling factor for the visual loss, which we set to 1.

\begin{figure}
    \centering
    \includegraphics[width=\linewidth]{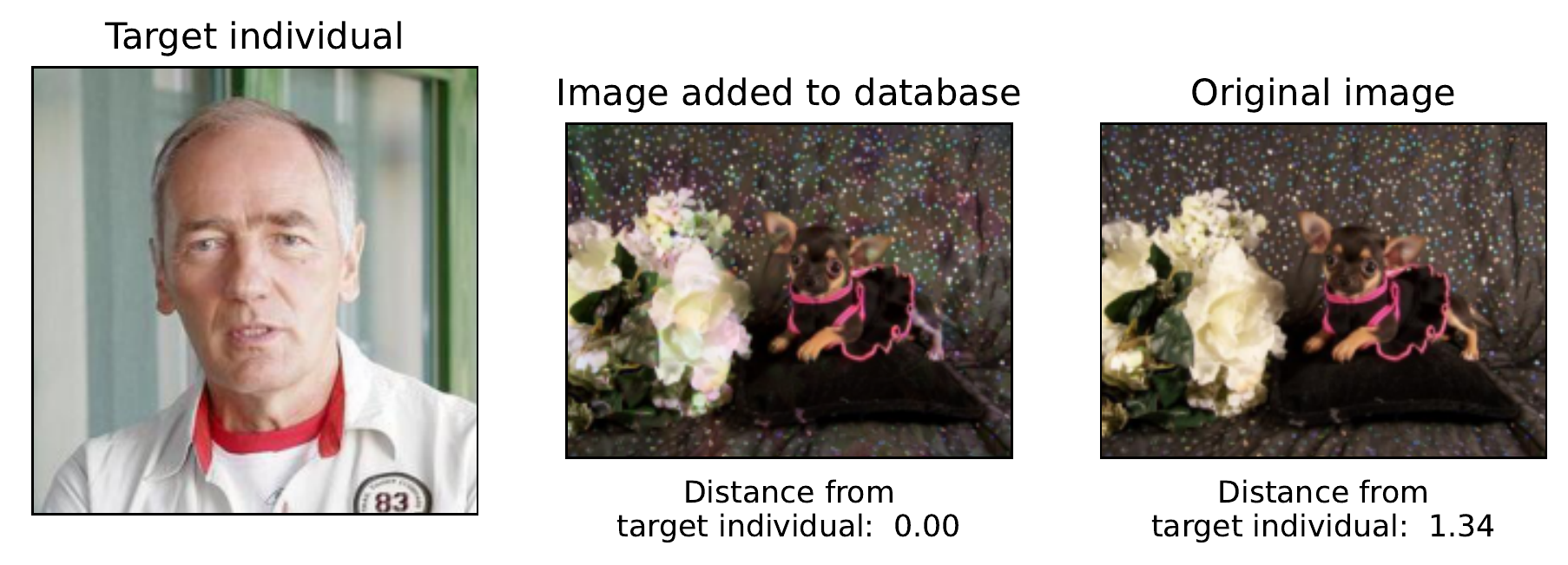}
    \vspace{-2em}
    \caption{Result of collision attack. The image on the right is modified so that its hash is close to one of the images of the target individual while remaining visually similar to the original image, which we here suppose to be illegal.}
    \label{fig:fr-collision}
\end{figure}

Fig.~\ref{fig:fr-collision} shows an example of the result of the collision attack for a target individual. 
The modified illegal image is visually similar to the original image while its hash is equal to the template hash. 
We obtain similar results for the template hash for all 10 targets with a median visual similarity of 0.98 $\pm$ 0.01 measured using structural similarity index measure (SSIM)~\cite{wang2004image}.
SSIM is a commonly used metric that uses distortion in structural information to measure the similarity between two images. SSIM $\in [0, 1]$, with larger values indicating higher similarity.
We evaluate the facial recognition performance when matching the images to the hash of modified illegal image. We obtain the same results as in our original experiment (Fig~\ref{fig:fr-target-performance}), an $F_1$ score of 51.2 $\pm$ 35 and recall of 67.2 $\pm$ 18.3, showing that indeed the hash of the modified illegal image is very similar to the template hash.

\textbf{Impact of increase in false positives (FP).} Table~\ref{tab:fr-performance-target} suggests that the number of FPs per million reported by dual-purpose (DP) models might be larger than for single-purpose (SP) models. If a CSS system flags too many FPs, this might lead auditors or companies compelled to install this algorithm to raise concerns. 
To answer this question, we however need to consider the number of FPs reported in the complete CSS system as this is what might be available to an auditor or the company, e.g. through mandatory reporting of flagged images.

The complete CSS system is composed of a PH algorithm $H$, reference set $R$, and threshold $T$. 
The single-purpose CSS system consists of an SP model $H_s$, a reference set of illegal images $R$, and a corresponding threshold $T_s$, while the dual-purpose CSS system would consist of a DP model $H_d$, reference set $R_d = R \cup R'$ where $R' = X^{I_T}_{\text{train}}$, and corresponding threshold $T_d$.

The number of FPs per million being flagged is, as expected, strongly influenced by the size of the reference set $R$, increasing with the database size~\cite{jain2022adversarial}. The reference database in the field, NCMEC, has been reported to contain more than 5 million unique hashes of illegal content~\cite{ncmec5million}. We do not have access to 5 million images but, to replicate real-world scenario as closely as possible, we here assume $R$ to be $R^{\text{primary}}$, i.e., the \textit{DISC21 reference set} and use Imagenet~\cite{russakovsky2015imagenet} as the set of non-illegal non-target query images the system would observe. Images from train, validation, and test sets of Imagenet are combined resulting in 1,431,093 images with $\sim17\%$ of images containing faces~\cite{yang2022study}. We evaluate the number of FPs flagged by SP and DP models against both reference sets $R^{\text{primary}}$, the single-purpose case, and $R_d = R^{\text{primary}} \cup R'$, the dual-purpose case. We have 10 sets of $R'$, i.e., $R'_i$, and thus 10 sets of $R_{d, i}$, corresponding each target individual $I_{T,i}$. False positives are evaluated for 10 single-purpose and 10 dual-purpose models on $R$. Each single-purpose model is evaluated on all 10 dual-purpose databases $R_{d,i}$, while the dual-purpose model for target $I_{T,i}$ is evaluated only against its corresponding reference $R_{d,i}$.

\begin{table}
    \caption{False-positives flagged in client-side scanning setup}\label{tab:fp-css-evaluation}
    \begin{center}
        \begin{tabular}{p{3.1cm}cc}
            \toprule
            \textbf{Models ($H$)} & \textbf{Single-purpose} & \textbf{Dual-purpose}\\
            & \textbf{Reference set $R$} & \textbf{Reference set $R_d$}\\
            \midrule
            Single-purpose models $H_s$ & 2873.2 $\pm$ 262.0 & 2873.2 $\pm$ 330.5\\
             \midrule
            Dual-purpose models $H_d$ & 2774.0 $\pm$ 575.2 & 2774.3 $\pm$ 575.2\\
            \bottomrule
        \end{tabular}
    \end{center}
    \vspace{-1.5em}
\end{table}

Table~\ref{tab:fp-css-evaluation} shows that the number of FPs flagged per million images by single-purpose and dual-purpose models each against their reference sets ($R$ for single purpose and $R_d$ for dual purpose) are extremely similar: $2873.2 \pm 262.0$ for the single purpose and $2774.3 \pm 575.2$ for the dual purpose. 
This shows that neither the change of model nor the addition of the templates to the database visibly impacts the number of FPs and therefore the total number of images reported. Our dual-purpose system would thus not generate more FPs, thereby not raising suspicions. 
We provide additional results on FPs with different thresholds and setups in Appendix~\ref{apd:fp-evaluation}. Finally, we note that the acceptable number of FPs in the context of client-side scanning is an open question that would mostly depend on what is acceptable from a political perspective (e.g. the current UK Online Safety Bill~\cite{ukonlineharms}).

\textbf{Hidden dual-purpose}.
We argue that the secondary purpose of the model would be hidden to the users and even to an auditor with access to our dual-purpose model and database of hashes (but not to the training dataset containing CSAM images), for two reasons. 
First, we demonstrate that our dual-purpose models are neither general facial recognition algorithms nor general face detectors, i.e. they do not flag every image with a face but are instead able to correctly identify the target individual in previously unknown pictures.
Second, we do not observe a noticeable difference between the activations of internal layers computed over images with and without faces.
Appendix~\ref{apd:activations-vis} shows that the internal layer activations of the dual-purpose model are visually similar for query images used for evaluating primary task performance and query images of non-target individuals used for evaluating secondary task performance. 
An auditor inspecting the model would thus not suspect that the dual-purpose model does something more than its advertised primary purpose of image copy detection.
Importantly though, we do not claim that the secondary purpose would be \textit{undetectable}. Indeed trivially, an auditor with access to images of the target individual will observe that our dual-purpose client-side scanning system indeed flags images of the target individual with a much larger probability than other individuals. This, however, assumes an auditor who knows what to look for, who to look for, and access to images of the target individual(s).

\textbf{Impact of LSH}. Embeddings from the models are converted into hashes using locality-sensitive hashing (LSH)~\cite{gionis1999similarity, applecsam}. We use Gaussian random projections followed by the Heaviside function as LSH to convert embeddings of size 256 to 256-bit hash. Applying LSH leads to a few single-purpose models to map many images to the same hash, strongly decreasing their ${\mu}AP$. As this only impacts the single-purpose model and would lead to unfair comparison, we exclude them for this experiment. Interestingly, no dual-purpose models suffered from this, possibly because of the dual-loss function we use.

\begin{table}
    \caption{Performance of models with LSH on image copy detection task}\label{tab:icd-performance-lsh}
    \begin{center}
        \begin{tabular}{ccc}
            \toprule
            \textbf{Metrics} & \textbf{Single-purpose} & \textbf{Dual-purpose}\\
            \midrule
            $\mu{AP}$ & 48.5 $\pm$ 0.8 & 48.4 $\pm$ 0.9\\
            \midrule
            Precision & 91.4 $\pm$ 0.8 & 91.1 $\pm$ 1.3\\
            \midrule
            Recall & 42.5 $\pm$ 0.8 & 42.5 $\pm$ 0.5\\
            \bottomrule
        \end{tabular}
    \end{center}
    \vspace{-1.5em}
\end{table}

\begin{table}
    \caption{Performance of models with LSH on facial recognition }\label{tab:fr-performance-lsh}
    \vspace{-1em}
    \begin{center}
        \begin{tabular}{ccc|cc}
            \toprule
            \multirow{2}*{\textbf{Metrics}} & \multicolumn{2}{c|}{\textbf{Single-purpose}} & \multicolumn{2}{c}{\textbf{Dual-purpose}}\\
            & \textbf{Target} & \textbf{Non-target} & \textbf{Target} & \textbf{Non-target} \\
            \midrule
            Recall & 0.6 $\pm$ 1.7 & 0.6 $\pm$ 1.5 & 66.2 $\pm$ 19.4 & 0.0 $\pm$ 0.7\\
            \midrule
            FP per M & 33.0 $\pm$ 65.7 & 29.2 $\pm$ 53.9 & 73.5 $\pm$ 45.2 & 3.0 $\pm$ 3.7\\
            \midrule
            Precision & 0.6 $\pm$ 2.4 & 1.6 $\pm$ 4.7 & 50.8 $\pm$ 39.6 & 0 $\pm$ 23.8\\
            \midrule
            $F_1$ score & 0.8 $\pm$ 1.3 & 0.9 $\pm$ 1.9 & 52.3 $\pm$ 34.5 & 0.0 $\pm$ 1.4\\
            \bottomrule
        \end{tabular}
    \end{center}
    \vspace{-1.5em}
\end{table}

Tables~\ref{tab:icd-performance-lsh} and~\ref{tab:fr-performance-lsh} shows the median performance of our models with LSH, on tasks of image copy detection and facial recognition respectively. As expected, ${\mu}AP$  decreases slightly, from 0.59 to 0.48 for both models, when LSH is applied. This is expected as LSH converts a vector of 256 floats to a binary hash of 256 bits leading to a significant loss of information. LSH however doesn't impact the ability of our dual-purpose models to identify target individuals. It achieves an $F_1$ score of 52.3, compared to 51.2 without LSH, and is still not a general facial recognition algorithm (low $F_1$ on non-targets individuals). This shows LSH does not impact the capacity of dual-purpose models to perform facial recognition.

\textbf{Impact of threshold $T$}. We have so far assumed the threshold $T$ is selected based on validation performance where the model achieves a precision of 90\% on the image copy detection task. Instead, a lower threshold could be chosen e.g. to limit the number of FPs. 

We here evaluate the model performance in test setup at thresholds corresponding to precision of 95\% and 99\% on the validation set for the primary task. Note that we do not report $\mu{AP}$ for the image copy detection task in Table~\ref{tab:performance-threshold-variation} as $\mu{AP}$ is independent of the chosen threshold and, hence, remains the same for all the models.

Table~\ref{tab:performance-threshold-variation} shows that all our results hold even when a lower threshold is selected. 
The single-purpose and dual-purpose models have similar performance on image copy detection task ($ICD$) and the dual-purpose models continue to reliably detect and match images of target-individuals ($F_1$-score for facial recognition on target individuals ($FR_{T}$)).

\begin{table*}
    \caption{Performance of models with different thresholds on  image copy detection ($ICD$) and facial recognition ($FR$) tasks}\label{tab:performance-threshold-variation}
    \begin{center}
        \begin{tabular}{ccc|c|c|c}
            \toprule
            \multirow{3}*{\textbf{Task}} & \multirow{3}*{\textbf{Metrics}} & \multicolumn{4}{c}{\textbf{Threshold ($T$)}}\\
            & & \multicolumn{2}{c|}{$T@95$} & \multicolumn{2}{c}{$T@99$} \\
            & & \textbf{Single-purpose} & \textbf{Dual-purpose} & \textbf{Single-purpose} & \textbf{Dual-purpose} \\
            \midrule
            \multirow{2}*{$ICD$} & Precision & 95.8 $\pm$ 0.5 & 95.4 $\pm$ 0.8 & 99.0 $\pm$ 0.2 & 99.0 $\pm$ 0.3\\
            \cmidrule{2-6}
            & Recall & 43.8 $\pm$ 1.3 & 44.0 $\pm$ 1.7 & 32.8 $\pm$ 1.5 & 32.3 $\pm$ 1.9\\
            \midrule
            \multirow{4}*{$FR_{T}$} & Recall & 0 $\pm$ 0.6 & 66.8 $\pm$ 18.8 & 0 $\pm$ 0 & 65.5 $\pm$ 18.5\\
            \cmidrule{2-6}
            & FP per million & 0 $\pm$ 0.6 & 76.1 $\pm$ 46.3 & 0 $\pm$ 0 & 66.4 $\pm$ 40.1\\
            \cmidrule{2-6}
            & Precision & 0 $\pm$ 7.9 & 49.8 $\pm$ 38.3 & 0 $\pm$ 0 & 52.9 $\pm$ 39.1\\
            \cmidrule{2-6}
            & $F_1$ score & 0 $\pm$ 1.2 & 52.0 $\pm$ 34.0 & 0 $\pm$ 0 & 53.4 $\pm$ 32.9\\
            \midrule
            \multirow{4}*{$FR_{T'}$} & Recall & 0 $\pm$ 0.5 & 0 $\pm$ 0 & 0 $\pm$ 0 & 0 $\pm$ 0\\
            \cmidrule{2-6}
            & FP per million & 0 $\pm$ 0.6 & 0 $\pm$ 0 & 0 $\pm$ 0 & 0 $\pm$ 0\\
            \cmidrule{2-6}
            & Precision & 0 $\pm$ 50.0 & 0 $\pm$ 0 & 0 $\pm$ 0 & 0 $\pm$ 0\\
            \cmidrule{2-6}
            & $F_1$ score & 0 $\pm$ 1.1 & 0 $\pm$ 0 & 0 $\pm$ 0 & 0 $\pm$ 0\\
            \bottomrule
        \end{tabular}
    \end{center}
    \vspace{-1em}
\end{table*}

\begin{table}
    \centering
    \caption{Facial recognition performance on target individuals when trained with $N^T_{\text{train}}$ images}\label{tab:fr-target-n-train}
    \begin{tabular}{ccccc}
        \toprule
        $N^T_{\text{train}}$ & \textbf{Recall} & \textbf{FP per M} & \textbf{Precision} & \textbf{$F_1$ score}\\
        \midrule
        100 & 67.2 $\pm$ 18.3 & 82.7 $\pm$ 51.8 & 48.0 $\pm$ 38.3 & 51.2 $\pm$ 35.0\\
        \midrule
        75 & 62.3 $\pm$ 9.1 & 94.5 $\pm$ 75.3 & 47.9 $\pm$ 37.9 & 51.9 $\pm$ 30.9\\
        \midrule
        50 & 54.3 $\pm$ 21.2 & 53.5 $\pm$ 48.8 & 48.5 $\pm$ 49.5 & 44.4 $\pm$ 31.8\\
        \midrule
        25 & 34.7 $\pm$ 13.5 & 65.3 $\pm$ 98.0 & 29.6 $\pm$ 54.6 & 33.4 $\pm$ 33.0\\
        \bottomrule
    \end{tabular}
\end{table}

\textbf{Impact of number of images in training}. Our dual-purpose models are trained using $N^T_{\text{train}}$ images of target individual $I_T$. In practice, for highly secretive individuals, the attacker might not have 100 images available. We here explore the impact of $N^T_{\text{train}}$ on the facial recognition performance of dual-purpose models on the target individual. To do so, we train the dual-purpose models with the same parameters as before except for the number of training samples available for the target individual. We select the best model with the same strategy and evaluate it in the test setup using the $F_1$ score. For evaluation, we use the same query set $Q^{\text{secondary}}_{\text{test}}$ as defined in Section~\ref{sec:fr-performance}.

Table~\ref{tab:fr-target-n-train} shows that decreasing the number of training images $N^T_{\text{train}}$ from 100 to 75 and even 50 training images only slightly impact the capabilities of the model. When very few images are available, less than 50, the $F_1$-score decreases sharply. Fewer images of target individual $I_T$ in training indeed reduces the diversity of images available for the model to learn from, preventing, in some cases, the model from learning a robust representation of the target individual. More advanced techniques might however be available in such cases~\cite{tan2006face}.

\begin{table}
    \caption{facial recognition performance of dual-purpose model trained to identify all 10 targets}\label{tab:fr-performance-all-10}
    \begin{center}
        \begin{tabular}{ccc}
            \toprule
            \textbf{Metrics} & \textbf{Target} & \textbf{Non-target}    \\
            \midrule
            Recall      & 51.3 $\pm$ 28.8  & 0.3 $\pm$ 0.9 \\
            \midrule
            FP per million    & 97 $\pm$ 95     & 0 $\pm$ 1     \\
            \midrule
            Precision   & 42.1 $\pm$ 53.6 & 0.4 $\pm$ 100.0 \\
            \midrule
            $F_1$ score & 46.0 $\pm$ 30.6  & 0.5 $\pm$ 1.6 \\
            \bottomrule
        \end{tabular}
    \end{center}
    \vspace{-1.5em}
\end{table}

\textbf{Single-model for all 10 targets}. So far our dual-purpose models have focused on identifying 1 target individual $I_T$ per model. In practice, one might want a model to identify several people e.g. the FBI 10 most wanted fugitives~\cite{fbitopten}. 

We evaluate the ability of our model to search for multiple people at once by training a dual-purpose model to identify all 10 target individuals with the same model. We use the same procedure and hyperparameters as used for dual-purpose models for 1 target. For the training batch $B_{\text{secondary}}$, we now sample an image from every target individual $I_{T,i}$ with probability $p_T$. Images from non-target individuals are thus sampled with probability $1 - 10 * p_T$. We evaluate the performance of the resulting dual-purpose model on image copy detection task, on facial recognition for all 10 targets, and non-target individuals. 

Table~\ref{tab:fr-performance-all-10} shows that aiming for the model to be able to perform facial recognition on 10 people at the same time is possible although at a slight cost in terms of its performance. The median $F_1$-score decreases from 51.2\% to 46\% while recall, the ability of the model to flag a new unknown picture of a target individual, decreases from 67.2\% to 51.3\%. Importantly though, it does not impact its performance on the primary task, $\mu{AP}$ of 59.7, nor does it turn it into a general facial recognition algorithm, $F_1$ of 0.5 on non-target individuals. We hypothesize that this slight decrease in the facial recognition performance of target individuals is due to the fairly limited capacity of the model and the relative simplicity of the training procedure. More complex model might have the ability to encode facial recognition capabilities for more targets at once and more sophisticated training procedure might help encode information about more individuals even in small models. 

\textbf{Implications.} Our research shows that a PH-based client-side scanning system can be designed to provide state-of-the-art performance on the primary task of image copy detection while also having a hidden secondary purpose of identifying target individual(s). We here exploit the typical expressivity and overparametrization of deep learning with a novel training strategy to learn the secondary task without impacting the performance on the primary task. Our results could likely be further improved with the extensive computational power available to a government attacker, e.g. to perform an extensive search over hyperparameters to train models with improved results on the secondary task. Our results are therefore only a lower bound on the performance a dual-purpose deep perceptual hashing algorithm can achieve. Taken together our results raises serious concerns that client-side scanning systems could be deployed with a ``hidden" feature of identifying target individuals, thus using billions of user devices as surveillance tools.

\section{Related work}\label{sec:related-work}

\textbf{Attacks on perceptual hashing-based client-side scanning (PH-CSS).} 
PH-CSS systems have repeatedly been shown to be vulnerable to both detection avoidance (evasion) and collision attacks. 
In \textit{detection avoidance} (DA) attacks, the goal of the attacker is to modify an image with an imperceptible perturbation to avoid detection~\cite{jain2022adversarial,wong2022blogpost,struppek2022facct,prokossquint, hao2021s}. DA attacks would allow a perpetrator to share illegal images without being detected.
In \textit{collision attacks}, images could be imperceptibly modified so that the hash of the modified image is similar to the hash of an illegal image in the database~\cite{dolhansky2020adversarial,struppek2022facct,prokossquint}. 
Collision attacks could be used to get an innocent account flagged.
Abelson et al.\cite{abelson2021bugs} provide a detailed overview of PH-CSS and existing concerns. 
In this paper, we introduce a new unknown vulnerability. More specifically we (1) demonstrate the perceptual hashing algorithm inherently does not have facial recognition capabilities, (2) formalize a realistic threat model where a perceptual hashing algorithm has a hidden feature of facial recognition of target individuals, (3) describe explicitly how the hidden feature could be added, through a specialized training procedure, to the algorithm, without building a general facial recognition capability into it, and (4) demonstrate its effectiveness.

\textbf{Dataset poisoning and backdoor attacks.} In dataset poisoning, the aim of the attacker is to manipulate the training data with an intention to cause models to fail during inference~\cite{shafahi2018poison,schwarzschild2021just}. Dataset poisoning is commonly used in backdoor attacks where the aim of the attacker is to hide an attacker-specific trigger in the model. When an input is presented to the model with the trigger, the model behaves in a predefined way for those inputs, e.g. misclassify the input to a specific class~\cite{li2022backdoor, liu2020reflection, saha2020hidden, chen2017targeted}. While dataset poisoning and backdoor attacks are designed for model to fail at inference times, e.g. misclassifying new data, our attack aims to build a secondary hidden feature into a model.

\textbf{Other perceptual hashing algorithms.} In this work, we add a secondary feature to the Yokoo model~\cite{yokoo2021contrastive}, a deep perceptual hashing (PH) algorithm trained with contrastive loss and cross-batch memory~\cite{wang2020cross}. Yokoo model won the ISC 2021 challenge to build models for the image copy detection task~\cite{papakipos2022results}. Other deep PH algorithms exist~\cite{pizzi2022self, wang2021bag, papadakis2021producing}, including SSCD~\cite{pizzi2022self} which uses SimCLR loss~\cite{chen2020simple}, a variant for contrastive learning and other models. In our setup the attacker has the ability to select the model and training procedure. We believe our results could be extended to other models, training procedures, and loss functions. Apart from the deep PH algorithms, there exists other non-learning based (so-called shallow) PH algorithms, like PDQ~\cite{facebookpdq}, and pHash~\cite{Zauner_2010}, that create the hash by extracting images features using predefined feature extractors like discrete cosine transforms~\cite{Ahmed_Natarajan_Rao_1974}. We believe that embedding a ``hidden" secondary purpose would be more difficult in a shallow hashing algorithm but not impossible. This would indeed e.g. require an attacker to train a model using a procedure like ours before ``reducing'' it to a shallow algorithm e.g. through pruning. Deep PH algorithms have furthermore been strongly outperforming shallow hashing algorithms, e.g. in Facebook's Image Search Challenge~\cite{papakipos2022results}. This means that while switching from deep PH to shallow PH might make the attack more difficult, this will come at a cost to the performance on primary task and potentially millions more false positive images being decrypted.

\bibliographystyle{IEEEtranS}
\bibliography{references}

\begin{thebibliography}{10}
\providecommand{\url}[1]{#1}
\csname url@samestyle\endcsname
\providecommand{\newblock}{\relax}
\providecommand{\bibinfo}[2]{#2}
\providecommand{\BIBentrySTDinterwordspacing}{\spaceskip=0pt\relax}
\providecommand{\BIBentryALTinterwordstretchfactor}{4}
\providecommand{\BIBentryALTinterwordspacing}{\spaceskip=\fontdimen2\font plus
\BIBentryALTinterwordstretchfactor\fontdimen3\font minus
  \fontdimen4\font\relax}
\providecommand{\BIBforeignlanguage}[2]{{%
\expandafter\ifx\csname l@#1\endcsname\relax
\typeout{** WARNING: IEEEtranS.bst: No hyphenation pattern has been}%
\typeout{** loaded for the language `#1'. Using the pattern for}%
\typeout{** the default language instead.}%
\else
\language=\csname l@#1\endcsname
\fi
#2}}
\providecommand{\BIBdecl}{\relax}
\BIBdecl

\bibitem{whatsappe2ee}
``About end-to-end encryption,''
  \url{https://faq.whatsapp.com/820124435853543/}, accessed on Dec 3, 2022.

\bibitem{bumbledetector}
\BIBentryALTinterwordspacing
``\BIBforeignlanguage{en}{Bumble releases open-source version of private
  detector a.i. feature to help tech platforms combat cyberflashing}.''
  [Online]. Available:
  \url{https://bumble.com/the-buzz/bumble-open-source-private-detector-ai-cyberflashing-dick-pics}
\BIBentrySTDinterwordspacing

\bibitem{uscsealaw}
``Certain activities relating to material involving the sexual exploitation of
  minors,'' \url{https://www.law.cornell.edu/uscode/text/18/2252}, accessed on
  Sep 21, 2021.

\bibitem{applecsam}
\BIBentryALTinterwordspacing
``Csam detection - technical summary 2021.'' [Online]. Available:
  \url{https://www.apple.com/child-safety/pdf/CSAM_Detection_Technical_Summary.pdf}
\BIBentrySTDinterwordspacing

\bibitem{ncmec5million}
\BIBentryALTinterwordspacing
``\BIBforeignlanguage{en}{Cybertipline data}.'' [Online]. Available:
  \url{http://www.missingkids.org/gethelpnow/cybertipline/cybertiplinedata.html}
\BIBentrySTDinterwordspacing

\bibitem{icloud}
\BIBentryALTinterwordspacing
``\BIBforeignlanguage{en}{icloud security overview}.'' [Online]. Available:
  \url{https://support.apple.com/en-us/HT202303}
\BIBentrySTDinterwordspacing

\bibitem{pie2ee}
\BIBentryALTinterwordspacing
``Securing privacy: Pi on end-to-end encryption | privacy international.''
  [Online]. Available:
  \url{https://privacyinternational.org/report/4949/securing-privacy-end-end-encryption}
\BIBentrySTDinterwordspacing

\bibitem{signal}
\BIBentryALTinterwordspacing
``\BIBforeignlanguage{en}{Signal messenger}.'' [Online]. Available:
  \url{https://signal.org/}
\BIBentrySTDinterwordspacing

\bibitem{fbitopten}
\BIBentryALTinterwordspacing
``Ten most wanted fugitives | fbi.'' [Online]. Available:
  \url{https://www.fbi.gov/wanted/topten}
\BIBentrySTDinterwordspacing

\bibitem{photodna}
\BIBentryALTinterwordspacing
``\BIBforeignlanguage{en-us}{Photodna | microsoft},'' 2009. [Online].
  Available: \url{https://www.microsoft.com/en-us/photodna}
\BIBentrySTDinterwordspacing

\bibitem{rsa2014}
\BIBentryALTinterwordspacing
``\BIBforeignlanguage{en}{How the nsa (may have) put a backdoor in rsa’s
  cryptography},'' Jan 2014. [Online]. Available:
  \url{http://blog.cloudflare.com/how-the-nsa-may-have-put-a-backdoor-in-rsas-cryptography-a-technical-primer/}
\BIBentrySTDinterwordspacing

\bibitem{cryptowars}
\BIBentryALTinterwordspacing
``\BIBforeignlanguage{en}{Crypto wars},'' Dec 2022, page Version ID:
  1130147393. [Online]. Available:
  \url{https://en.wikipedia.org/w/index.php?title=Crypto_Wars&oldid=1130147393}
\BIBentrySTDinterwordspacing

\bibitem{ofcomoverview}
\BIBentryALTinterwordspacing
``\BIBforeignlanguage{en}{Overview of perceptual hashing technology},'' Nov
  2022. [Online]. Available:
  \url{https://www.ofcom.org.uk/research-and-data/online-research/overview-of-perceptual-hashing-technology}
\BIBentrySTDinterwordspacing

\bibitem{euregulationcsam}
\BIBentryALTinterwordspacing
\emph{\BIBforeignlanguage{en}{Proposal for a REGULATION OF THE EUROPEAN
  PARLIAMENT AND OF THE COUNCIL laying down rules to prevent and combat child
  sexual abuse}}, 2022. [Online]. Available:
  \url{https://eur-lex.europa.eu/legal-content/EN/TXT/?uri=COM\%3A2022\%3A209\%3AFIN&qid=1652451192472}
\BIBentrySTDinterwordspacing

\bibitem{abelson2021bugs}
H.~Abelson, R.~Anderson, S.~M. Bellovin, J.~Benaloh, M.~Blaze, J.~Callas,
  W.~Diffie, S.~Landau, P.~G. Neumann, R.~L. Rivest \emph{et~al.}, ``Bugs in
  our pockets: The risks of client-side scanning,'' \emph{arXiv preprint
  arXiv:2110.07450}, 2021.

\bibitem{Ahmed_Natarajan_Rao_1974}
N.~Ahmed, T.~Natarajan, and K.~R. Rao, ``Discrete cosine transform,''
  \emph{IEEE TRANSACTIONS ON COMPUTERS}, p.~4, 1974.

\bibitem{babenko2014neural}
A.~Babenko, A.~Slesarev, A.~Chigorin, and V.~Lempitsky, ``Neural codes for
  image retrieval,'' in \emph{European conference on computer vision}.\hskip
  1em plus 0.5em minus 0.4em\relax Springer, 2014, pp. 584--599.

\bibitem{biswas2021state}
R.~Biswas and P.~Blanco-Medina, ``State of the art: Image hashing,''
  \emph{arXiv preprint arXiv:2108.11794}, 2021.

\bibitem{blaze2011key}
M.~Blaze, ``Key escrow from a safe distance: looking back at the clipper
  chip,'' in \emph{Proceedings of the 27th Annual Computer Security
  Applications Conference}, 2011, pp. 317--321.

\bibitem{whatsappblog2017}
W.~blog, ``Connecting one billion users every day,''
  \url{https://blog.whatsapp.com/connecting-one-billion-users-every-day}, 2017,
  accessed on Dec 3, 2022.

\bibitem{whatsapp2billion}
W.~Blog, ``Two billion users -- connecting the world privately,''
  \url{https://blog.whatsapp.com/two-billion-users-connecting-the-world-privately/},
  2020, accessed on Dec 3, 2022.

\bibitem{cao2018vggface2}
Q.~Cao, L.~Shen, W.~Xie, O.~M. Parkhi, and A.~Zisserman, ``Vggface2: A dataset
  for recognising faces across pose and age,'' in \emph{2018 13th IEEE
  international conference on automatic face \& gesture recognition (FG
  2018)}.\hskip 1em plus 0.5em minus 0.4em\relax IEEE, 2018, pp. 67--74.

\bibitem{chen2020simple}
T.~Chen, S.~Kornblith, M.~Norouzi, and G.~Hinton, ``A simple framework for
  contrastive learning of visual representations,'' in \emph{International
  conference on machine learning}.\hskip 1em plus 0.5em minus 0.4em\relax PMLR,
  2020, pp. 1597--1607.

\bibitem{chen2017targeted}
X.~Chen, C.~Liu, B.~Li, K.~Lu, and D.~Song, ``Targeted backdoor attacks on deep
  learning systems using data poisoning,'' \emph{arXiv preprint
  arXiv:1712.05526}, 2017.

\bibitem{Deng2020CVPR}
J.~Deng, J.~Guo, E.~Ververas, I.~Kotsia, and S.~Zafeiriou, ``Retinaface:
  Single-shot multi-level face localisation in the wild,'' in \emph{CVPR},
  2020.

\bibitem{dolhansky2020adversarial}
B.~Dolhansky and C.~C. Ferrer, ``Adversarial collision attacks on image hashing
  functions,'' \emph{arXiv:2011.09473}, 2020.

\bibitem{douze20212021}
M.~Douze, G.~Tolias, E.~Pizzi, Z.~Papakipos, L.~Chanussot, F.~Radenovic,
  T.~Jenicek, M.~Maximov, L.~Leal-Taix{\'e}, I.~Elezi \emph{et~al.}, ``The 2021
  image similarity dataset and challenge,'' \emph{arXiv preprint
  arXiv:2106.09672}, 2021.

\bibitem{ukonlineharms}
D.~for DCMS, ``Draft online safety bill,''
  \url{https://www.gov.uk/government/publications/draft-online-safety-bill},
  accessed on Sep 21, 2021.

\bibitem{gionis1999similarity}
A.~Gionis, P.~Indyk, R.~Motwani \emph{et~al.}, ``Similarity search in high
  dimensions via hashing,'' in \emph{Vldb}, vol.~99, no.~6, 1999, pp. 518--529.

\bibitem{glorot2010understanding}
X.~Glorot and Y.~Bengio, ``Understanding the difficulty of training deep
  feedforward neural networks,'' in \emph{Proceedings of the thirteenth
  international conference on artificial intelligence and statistics}.\hskip
  1em plus 0.5em minus 0.4em\relax JMLR Workshop and Conference Proceedings,
  2010, pp. 249--256.

\bibitem{guo2021sample}
J.~Guo, J.~Deng, A.~Lattas, and S.~Zafeiriou, ``Sample and computation
  redistribution for efficient face detection,'' \emph{arXiv preprint
  arXiv:2105.04714}, 2021.

\bibitem{hao2021s}
Q.~Hao, L.~Luo, S.~T. Jan, and G.~Wang, ``It's not what it looks like:
  Manipulating perceptual hashing based applications,'' in \emph{Proceedings of
  the 2021 ACM SIGSAC Conference on Computer and Communications Security},
  2021, pp. 69--85.

\bibitem{Iglovikov_2020}
\BIBentryALTinterwordspacing
V.~Iglovikov, ``\BIBforeignlanguage{en}{Face recognition on 330 million images
  at 400 images per second},'' Aug 2020. [Online]. Available:
  \url{https://towardsdatascience.com/face-recognition-on-330-million-images-at-400-images-per-second-b85e594eab66}
\BIBentrySTDinterwordspacing

\bibitem{jain2022adversarial}
S.~Jain, A.-M. Crețu, and Y.-A. de~Montjoye, ``Adversarial detection avoidance
  attacks: Evaluating the robustness of perceptual hashing-based client-side
  scanning,'' in \emph{31st USENIX Security Symposium (USENIX Security 22)},
  2022, pp. 2317--2334.

\bibitem{johnson2019billion}
J.~Johnson, M.~Douze, and H.~J{\'e}gou, ``Billion-scale similarity search with
  {GPUs},'' \emph{IEEE Transactions on Big Data}, vol.~7, no.~3, pp. 535--547,
  2019.

\bibitem{facebookpdq}
J.~Kerl, ``{The TMK+PDQF video-hashing algorithim and the PDQ image hashing
  algorithm},''
  \url{https://github.com/facebook/ThreatExchange/blob/master/hashing/hashing.pdf},
  2020, accessed on June 7, 2021.

\bibitem{khosla2011novel}
A.~Khosla, N.~Jayadevaprakash, B.~Yao, and F.-F. Li, ``Novel dataset for
  fine-grained image categorization: Stanford dogs,'' in \emph{Proc. CVPR
  workshop on fine-grained visual categorization (FGVC)}, vol.~2.\hskip 1em
  plus 0.5em minus 0.4em\relax Citeseer, 2011.

\bibitem{kulshrestha2021identifying}
A.~Kulshrestha and J.~Mayer, ``Identifying harmful media in $\{$End-to-End$\}$
  encrypted communication: Efficient private membership computation,'' in
  \emph{30th USENIX Security Symposium (USENIX Security 21)}, 2021, pp.
  893--910.

\bibitem{principles2018key}
I.~Levy and C.~Robinson, ``Principles for a more informed exceptional access
  debate,''
  \url{https://www.lawfareblog.com/principles-more-informed-exceptional-access-debate},
  2018.

\bibitem{levy2022thoughts}
------, ``Thoughts on child safety on commodity platforms,'' \emph{arXiv
  preprint arXiv:2207.09506}, 2022.

\bibitem{li2022backdoor}
Y.~Li, Y.~Jiang, Z.~Li, and S.-T. Xia, ``Backdoor learning: A survey,''
  \emph{IEEE Transactions on Neural Networks and Learning Systems}, 2022.

\bibitem{liu2020reflection}
Y.~Liu, X.~Ma, J.~Bailey, and F.~Lu, ``Reflection backdoor: A natural backdoor
  attack on deep neural networks,'' in \emph{European Conference on Computer
  Vision}.\hskip 1em plus 0.5em minus 0.4em\relax Springer, 2020, pp. 182--199.

\bibitem{musgrave2020pytorch}
K.~Musgrave, S.~Belongie, and S.-N. Lim, ``Pytorch metric learning,'' 2020.

\bibitem{ncmec2019}
NCMEC, ``Ncmec’s statement regarding end-to-end encryption,''
  \url{https://www.missingkids.org/blog/2019/post-update/end-to-end-encryption},
  Mar 2019, accessed on June 6, 2021.

\bibitem{ukcsealaw}
H.~Office, ``Interim code of practice on online child sexual exploitation and
  abuse (accessible version),''
  \url{https://www.gov.uk/government/publications/online-harms-interim-codes-of-practice/interim-code-of-practice-on-online-child-sexual-exploitation-and-abuse-accessible-version},
  Dec 2020, accessed on June 6, 2021.

\bibitem{papadakis2021producing}
S.~M. Papadakis and S.~Addicam, ``Producing augmentation-invariant embeddings
  from real-life imagery,'' \emph{arXiv preprint arXiv:2112.03415}, 2021.

\bibitem{papakipos2022augly}
Z.~Papakipos and J.~Bitton, ``Augly: Data augmentations for robustness,'' 2022.

\bibitem{papakipos2022results}
Z.~Papakipos, G.~Tolias, T.~Jenicek, E.~Pizzi, S.~Yokoo, W.~Wang, Y.~Sun,
  W.~Zhang, Y.~Yang, S.~Addicam \emph{et~al.}, ``Results and findings of the
  2021 image similarity challenge,'' in \emph{NeurIPS 2021 Competitions and
  Demonstrations Track}.\hskip 1em plus 0.5em minus 0.4em\relax PMLR, 2022, pp.
  1--12.

\bibitem{paszke2019pytorch}
A.~Paszke, S.~Gross, F.~Massa, A.~Lerer, J.~Bradbury, G.~Chanan, T.~Killeen,
  Z.~Lin, N.~Gimelshein, L.~Antiga \emph{et~al.}, ``Pytorch: An imperative
  style, high-performance deep learning library,'' \emph{Advances in neural
  information processing systems}, vol.~32, 2019.

\bibitem{international2020e2ee}
P.~Patel, W.~Barr, P.~Dutton, A.~Little, B.~Blair, India, and Japan,
  ``International statement: End-to-end encryption and public safety,''
  \url{https://www.gov.uk/government/publications/international-statement-end-to-end-encryption-and-public-safety},
  Oct 2020, accessed on June 6, 2021.

\bibitem{pizzi2022self}
E.~Pizzi, S.~D. Roy, S.~N. Ravindra, P.~Goyal, and M.~Douze, ``A
  self-supervised descriptor for image copy detection,'' in \emph{Proceedings
  of the IEEE/CVF Conference on Computer Vision and Pattern Recognition}, 2022,
  pp. 14\,532--14\,542.

\bibitem{prokossquint}
J.~Prokos, N.~Fendley, M.~Green, R.~Schuster, E.~Tromer, T.~M. Jois, and
  Y.~Cao, ``Squint hard enough: Attacking perceptual hashing with adversarial
  machine learning.''

\bibitem{radenovic2018fine}
F.~Radenovi{\'c}, G.~Tolias, and O.~Chum, ``Fine-tuning cnn image retrieval
  with no human annotation,'' \emph{IEEE transactions on pattern analysis and
  machine intelligence}, vol.~41, no.~7, pp. 1655--1668, 2018.

\bibitem{ridnik2021imagenet}
T.~Ridnik, E.~Ben-Baruch, A.~Noy, and L.~Zelnik-Manor, ``Imagenet-21k
  pretraining for the masses,'' \emph{arXiv preprint arXiv:2104.10972}, 2021.

\bibitem{russakovsky2015imagenet}
O.~Russakovsky, J.~Deng, H.~Su, J.~Krause, S.~Satheesh, S.~Ma, Z.~Huang,
  A.~Karpathy, A.~Khosla, M.~Bernstein \emph{et~al.}, ``Imagenet large scale
  visual recognition challenge,'' \emph{International journal of computer
  vision}, vol. 115, no.~3, pp. 211--252, 2015.

\bibitem{saha2020hidden}
A.~Saha, A.~Subramanya, and H.~Pirsiavash, ``Hidden trigger backdoor attacks,''
  in \emph{Proceedings of the AAAI conference on artificial intelligence},
  vol.~34, no.~07, 2020, pp. 11\,957--11\,965.

\bibitem{schwarzschild2021just}
A.~Schwarzschild, M.~Goldblum, A.~Gupta, J.~P. Dickerson, and T.~Goldstein,
  ``Just how toxic is data poisoning? a unified benchmark for backdoor and data
  poisoning attacks,'' in \emph{International Conference on Machine
  Learning}.\hskip 1em plus 0.5em minus 0.4em\relax PMLR, 2021, pp. 9389--9398.

\bibitem{shafahi2018poison}
A.~Shafahi, W.~R. Huang, M.~Najibi, O.~Suciu, C.~Studer, T.~Dumitras, and
  T.~Goldstein, ``Poison frogs! targeted clean-label poisoning attacks on
  neural networks,'' \emph{Advances in neural information processing systems},
  vol.~31, 2018.

\bibitem{whatsapp100billion}
M.~Singh, ``Whatsapp is now delivering roughly 100 billion messages a day,''
  \url{https://social.techcrunch.com/2020/10/29/whatsapp-is-now-delivering-roughly-100-billion-messages-a-day/},
  Oct 2020, accessed on June 6, 2021.

\bibitem{struppek2022facct}
L.~Struppek, D.~Hintersdorf, D.~Neider, and K.~Kersting, ``Learning to break
  deep perceptual hashing: The use case neuralhash,'' in \emph{Proceedings of
  the ACM Conference on Fairness, Accountability, and Transparency (FAccT)},
  2022.

\bibitem{tan2021efficientnetv2}
M.~Tan and Q.~Le, ``Efficientnetv2: Smaller models and faster training,'' in
  \emph{International Conference on Machine Learning}.\hskip 1em plus 0.5em
  minus 0.4em\relax PMLR, 2021, pp. 10\,096--10\,106.

\bibitem{tan2006face}
X.~Tan, S.~Chen, Z.-H. Zhou, and F.~Zhang, ``Face recognition from a single
  image per person: A survey,'' \emph{Pattern recognition}, vol.~39, no.~9, pp.
  1725--1745, 2006.

\bibitem{tolias2015particular}
G.~Tolias, R.~Sicre, and H.~J{\'e}gou, ``Particular object retrieval with
  integral max-pooling of cnn activations,'' \emph{arXiv preprint
  arXiv:1511.05879}, 2015.

\bibitem{wang2021bag}
W.~Wang, W.~Zhang, Y.~Sun, and Y.~Yang, ``Bag of tricks and a strong baseline
  for image copy detection,'' \emph{arXiv preprint arXiv:2111.08004}, 2021.

\bibitem{wang2020cross}
X.~Wang, H.~Zhang, W.~Huang, and M.~R. Scott, ``Cross-batch memory for
  embedding learning,'' in \emph{Proceedings of the IEEE/CVF Conference on
  Computer Vision and Pattern Recognition}, 2020, pp. 6388--6397.

\bibitem{wang2004image}
Z.~Wang, A.~C. Bovik, H.~R. Sheikh, and E.~P. Simoncelli, ``Image quality
  assessment: from error visibility to structural similarity,'' \emph{IEEE
  transactions on image processing}, vol.~13, no.~4, pp. 600--612, 2004.

\bibitem{rw2019timm}
R.~Wightman, ``Pytorch image models,''
  \url{https://github.com/rwightman/pytorch-image-models}, 2019.

\bibitem{wong2022blogpost}
A.~Wong, S.~Jain, A.-M. Cretu, and Y.-A. de~Montjoye, ``Blogpost: Deep
  perceptual hashing is not robust to adversarial detection avoidance
  attacks,'' 2022.

\bibitem{yang2022study}
K.~Yang, J.~H. Yau, L.~Fei-Fei, J.~Deng, and O.~Russakovsky, ``A study of face
  obfuscation in imagenet,'' in \emph{International Conference on Machine
  Learning}.\hskip 1em plus 0.5em minus 0.4em\relax PMLR, 2022, pp.
  25\,313--25\,330.

\bibitem{yokoo2021contrastive}
S.~Yokoo, ``Contrastive learning with large memory bank and negative embedding
  subtraction for accurate copy detection,'' \emph{arXiv preprint
  arXiv:2112.04323}, 2021.

\bibitem{Zauner_2010}
C.~Zauner, ``Implementation and benchmarking of perceptual image hash
  functions,'' Master's thesis, 2010,
  \url{https://www.phash.org/docs/pubs/thesis_zauner.pdf}.

\bibitem{zhang2020method}
K.~Zhang, V.~Albiero, and K.~W. Bowyer, ``A method for curation of web-scraped
  face image datasets,'' in \emph{2020 8th International Workshop on Biometrics
  and Forensics (IWBF)}.\hskip 1em plus 0.5em minus 0.4em\relax IEEE, 2020, pp.
  1--6.

\bibitem{zhong2017random}
Z.~Zhong, L.~Zheng, G.~Kang, S.~Li, and Y.~Yang, ``Random erasing data
  augmentation. arxiv,'' \emph{arXiv preprint arXiv:1708.04896}, 2017.

\bibitem{zhu2021webface260m}
Z.~Zhu, G.~Huang, J.~Deng, Y.~Ye, J.~Huang, X.~Chen, J.~Zhu, T.~Yang, J.~Lu,
  D.~Du \emph{et~al.}, ``Webface260m: A benchmark unveiling the power of
  million-scale deep face recognition,'' in \emph{Proceedings of the IEEE/CVF
  Conference on Computer Vision and Pattern Recognition}, 2021, pp.
  10\,492--10\,502.

\end{thebibliography}

\appendices

\section{Model initialization}\label{apd:model-initialization}

The CNN backbone in the Yokoo model is initialized at the start of training with weights obtained by pretraining on ImageNet21K~\cite{ridnik2021imagenet}. The fully-connected layer is initialized with random weights based on \textit{Xavier initialization} using normal distribution~\cite{glorot2010understanding}.

\section{Augmentations}\label{apd:augmentation-appendix}

\textbf{Augmentations.} Each augmentation contains a series of image transformation steps, that are applied sequentially on an input image with some probability $p$. We use the same augmentations as used by the Yokoo model~\cite{yokoo2021contrastive}. We describe the list transformations in augmentations $A_{m}$ and $A_{h}$, along with their corresponding probability $p$ in detail in Appendix~\ref{apd:augmentation-appendix}. Transformations 3 to 14 in $A_{h}$ are shuffled every time before applying to an image. All other transformations are applied in the same order as stated in the table. We use uniform sampling, unless specified otherwise.

\begin{table*}[!ht]
    \begin{center}
        \caption{List of image transformations for $A_{h}$}\label{tab:hard-augmentations}
        \begin{tabular}{|c|p{3cm}|p{1cm}|p{3cm}|p{7cm}|}
            \hline    
            \textbf{Order} & \textbf{Image transformation} & $\mathbf{p}$ & \textbf{Parameters} & \textbf{Description}\\ 
            \hline
            1 & Rotation & 0.25 & $p_1$=0, $p_2$=180 & Randomly rotates an image with a value between $p_1$ and $p_2$.\\
            2 & Overlay Image and Resized Crop & 1.0 & $D_{tr}$, $p_1$=0.05, $p_2$=0.6, $p_3$=0.4, $p_4$=0.6, $p_5$=0.7, $p_6$=0.15, $p_7$=256 & With probability $p_1$, the input image and another image sampled at random from $D_{tr}$ are combined with one acting as background and other as overlay. Opacity of overlay image is sampled from $[p_2, 1.0]$, and size of the overlay image is sampled from $[p_3, p_4]$. The position of overlay is determined by the size of the overlay image. And the overlaid image is then randomly cropped to a scale of $[p_5, 1.0]$ and resized to size $p_7 \times p_7$. With probability $(1-p_1)$ the input image is simply randomly cropped to a scale between $[p_6, 1.0]$ and resized to size $p_7 \times p_7$.\\
            3 & ColorJitter & 1.0 & $p_1$=0.7, $p_2$=0.7, $p_3$=0.7, $p_4$=0.2 & Randomly change the brightness, contrast, saturation and hue of the image. Brightness, contrast and saturation are changed by a factor sampled from $[1-p_1, 1+p_1]$, $[1-p_2, 1+p_2]$, and $[1-p_3, 1+p_3]$ respectively. Hue is jittered by a factor of sampled from $[-p_4, p_4]$.\\
            4 & Pixelation & 0.25 & $p_1$=0.1, $p_2$=1.0 & Randomly pixelate the image by a factor sampled from $[p_1, p_2]$. Lower value implies more pixelated image.\\
            5 & Shuffle Pixels & 0.25 & $p_1$=0.1 & Shuffle $p_1$ proportion of pixels.\\
            6 & Encoding Quality & 0.25 & $p_1=[10, 20, 30, 50]$ & Encode the Image in JPEG with quality sampled from $p_1$.\\
            7 & Grayscale & 0.25 & None & Convert image to grayscale\\
            8 & Blur & 0.25 & $p_1$=0.0, $p_2$=10.0 & Randomly blur the image with a radius sampled from $[p_1, p_2]$. Larger the radius, blurrier the image.\\
            9 & Perspective & 0.25 & $p_1$=0.5 & Apply a random perspective transform on the image with distortion scale of $p_1$.\\
            10 & Horizontal Flip & 0.25 & None & Horizontally flip the image.\\
            11 & Vertical Flip & 0.25 & None & Vertically flip the image.\\
            12 & Overlay Text & 0.25 & $p_1$=0.1, $p_2$=0.3 & Overlays a random text on top of the image, of size between $[p_1, p_2]$ and random color.\\
            13 & Emoji Overlay & 0.25 & $p_1$=0.1, $p_2$=0.3 & Overlay a random emoji on top of the image, of size between $[p_1, p_2]$.\\
            14 & Edge Enhancement & 0.25 & None & Apply edge enhancement to the image, a filter is sampled from one of the two standard edge enhancement filters and used.\\
            15 & Converts to RGB & 1.0 & None & Convert the image to 3-channel RGB image.\\
            16 & Erasing & 0.25 & None & Randomly selects a rectangle region from the image and erases its pixels with random values~\cite{zhong2017random}.\\
            \hline
        \end{tabular}
    \end{center}
    \vspace{-1em}
\end{table*}

\section{Implementation details}\label{appendix:implementation-details}

Our implementation for training the model is adapted from the code provided by the Yokoo model~\cite{yokoo2021contrastive}. The models are implemented in PyTorch~\cite{paszke2019pytorch}. The Timm library~\cite{rw2019timm} is used to create the CNN backbone and to initialize it with the pretrained weights. The PyTorch metric learning~\cite{musgrave2020pytorch} is used to implement the loss functions. Torchvision and Augly~\cite{papakipos2022augly} are used to implement augmentations while FAISS is used in evaluation~\cite{johnson2019billion}. To ensure reproducibility, we use a single seed to initialize our model as well as to control the sampling and the ordering of the data.

\begin{table*}[!ht]
    \begin{center}
        \caption{List of image transformations for $A_{m}$}\label{tab:moderate-augmentations}
        \vspace{-0.5em}
        \begin{tabular}{|c|p{3cm}|p{1cm}|p{3cm}|p{7cm}|}
            \hline    
            \textbf{Order} & \textbf{Image transformation} & $\mathbf{p}$ & \textbf{Parameters} & \textbf{Description}\\ 
            \hline
            1 & Resized Crop & 1.0 & $p_1$=0.7, $p_2$=1.0, $p_3$=256 & Crop a random portion of image of scale sampled from $[p_1, p_2]$ and resize it to size of $p_3 \times p_3$.\\
            2 & Horizontal Flip & 0.5 & None & Horizontally flip the image.\\
            \hline
        \end{tabular}
    \end{center}
    \vspace{-1em}
\end{table*}

Table~\ref{tab:train-hyperparams} lists the hyperparameters used to train the models and their corresponding values. The hyperparameters that are specific to dual-purpose models are marked by ``NA'' in the single-purpose column. In each iteration, the single-purpose models processes only $b_{\text{primary}}=96$ images corresponding to the primary task, while the dual-purpose model processes $b_{\text{primary}} + b_{\text{secondary}}=108$ images, 96 images corresponding to the primary task and 12 additional images corresponding to the secondary task. We consequently increase the memory size $M$ for XBM from 20000 for the single-purpose model to 22500 for the dual-purpose model.

\begin{figure}[!h]
    \centering
    \includegraphics[width=0.9\linewidth]{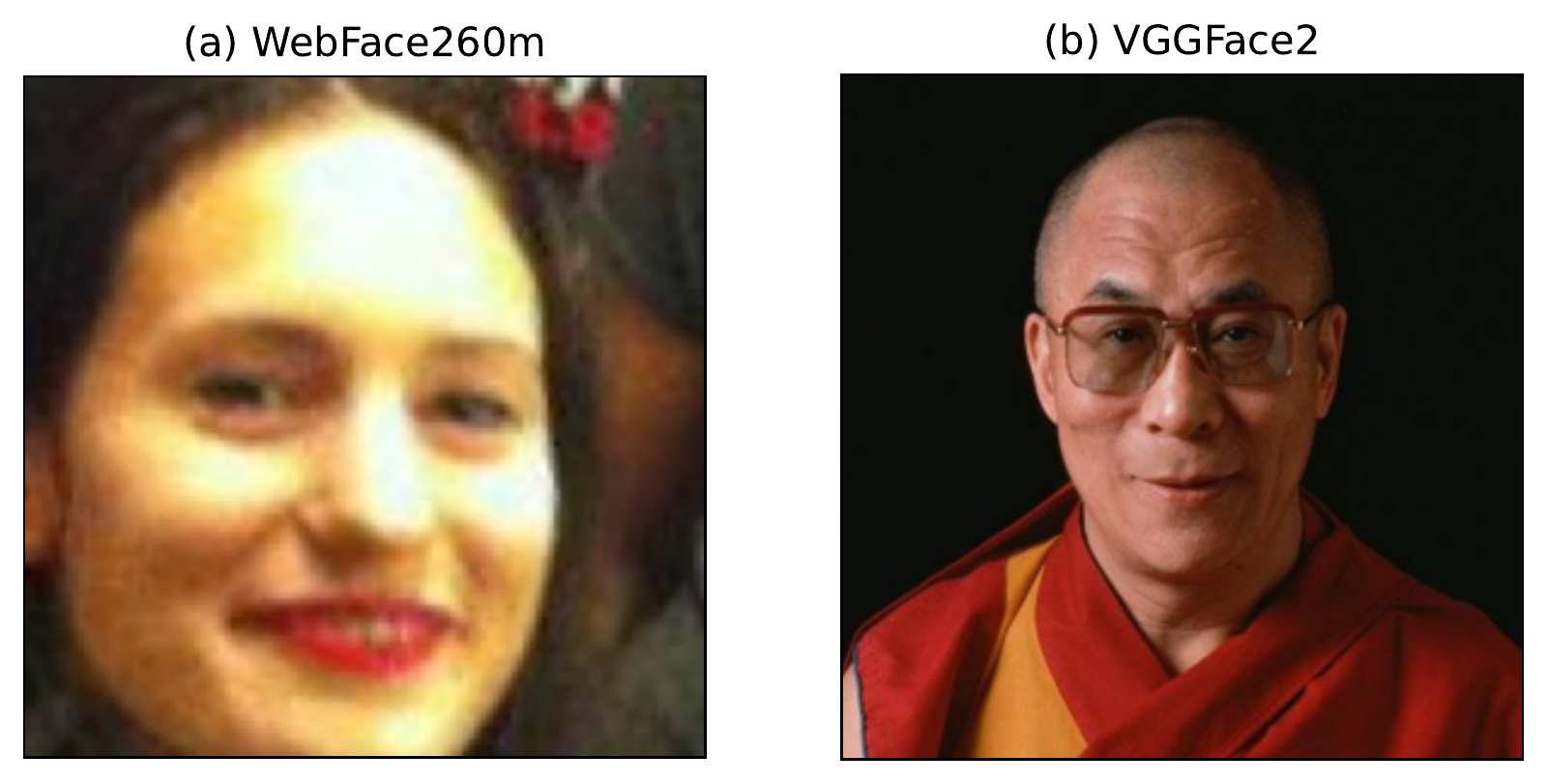}
    \caption{Examples of images from different facial recognition datasets. VGGFace2 dataset images (right) are more realistic than heavily cropped images from other facial recognition datasets like WebFace260m (left).}
    \label{fig:fr-dataset-images}
\end{figure}

\begin{figure*}[tbp]
    \begin{center}
        \includegraphics[width=0.8\linewidth]{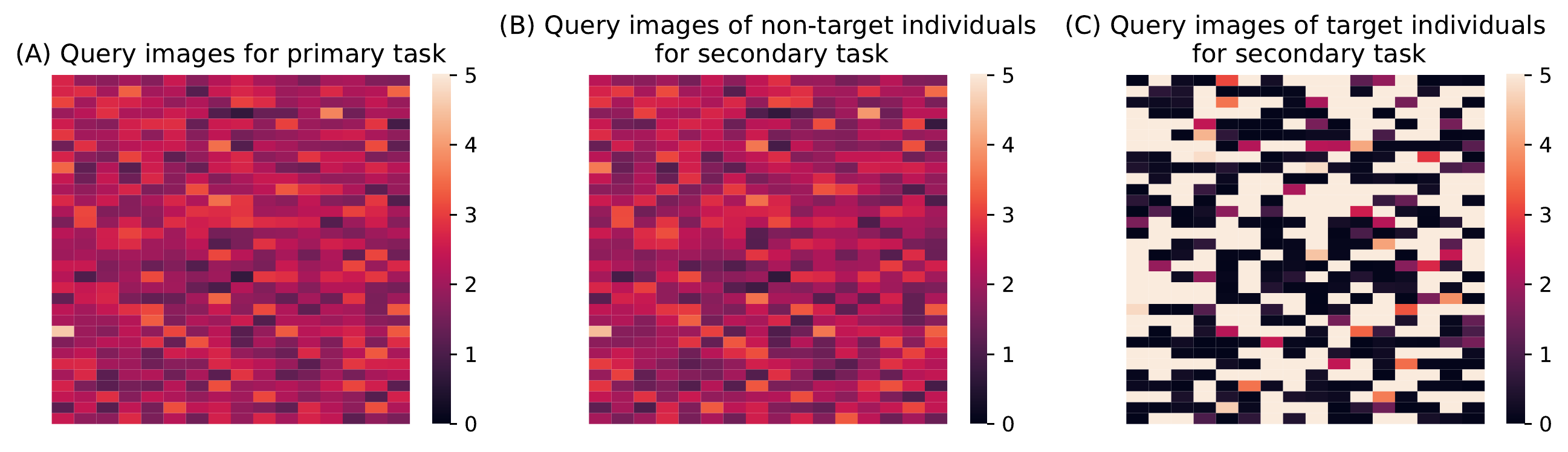}
        \caption{Heatmap of average activations of second to last layer for a dual-purpose model. For a better visualization, we reshape the 512 neuron activations to a 32$\times$16 matrix, and clip values to a maximum of 5.}
        \label{fig:activations}
    \end{center}
\end{figure*}

Our choice of hyperparameters is constrained by our computational resources and informed by the performances of the models on the validation set. All our models are trained on 2 GPUs with at least 54 cores and 108 GB of memory. It takes approximately 30 hours for 1 model to train. We are thus unable to perform exhaustive hyperparameter search. 
We also found the learning with $b_{\text{primary}}$ of 96 to be unstable for certain models and, due to limited compute resources, we were unable to increase the $b_{\text{primary}}$ beyond 96. We use gradient accumulation over 2 batches, i.e., we update the model after every 2 batches rather than 1. We do this by averaging gradients from both of the batches, which reduces the noise in the model update. Thus, the effective batch sizes for primary and secondary task are twice $b_{\text{primary}}$ and $b_{\text{secondary}}$. This strategy allowed us to train our models effectively even with limited computation resources.

\begin{table}[!ht]
    \caption{Hyperparameters for training}\label{tab:train-hyperparams}
    \begin{center}
        \begin{tabular}{|p{4cm}|c|p{0.9cm}|p{0.9cm}|}
            \hline
            \multirow{2}{*}{Parameter description} &  \multirow{2}{*}{Notation} & Single-purpose & Dual-purpose\\
            \hline
            \multicolumn{4}{|l|}{\textbf{SGD Optimizer parameters}}\\
            \hline
            Initial learning rate & & 0.1 & 0.1\\
            Learning rate decay & $\gamma$ & 0.9 & 0.9\\
            Minimum learning rate & $\eta_{min}$ & 0.05 & 0.05\\
            Weight decay rate & & 1e-6 & 1e-6\\
            Momentum & & 0.9 & 0.9\\
            \hline
            \multicolumn{4}{|l|}{\textbf{Loss function parameters}}\\
            \hline
            Margin for positive samples & $m_p$ & 0.0 & 0.0\\
            Margin for negative samples & $m_n$ & 1.0 & 1.0\\
            Memory size for XBM & $M$ & 20000 & 22500\\
            \hline
            \multicolumn{4}{|l|}{\textbf{Batching parameters}}\\
            \hline
            Batch size for primary task & $b_{\text{primary}}$ & 96 & 96\\
            Batch size for secondary task & $b_{\text{secondary}}$ & NA & 12\\
            Probability to sample image from target individual in $B_{\text{secondary}}$ & $p_T$ & NA & 0.025\\
            \hline
            \multicolumn{4}{|l|}{\textbf{Dual-purpose training parameters}}\\
            \hline
            Number of images of target individual for training & $N^T_{\text{train}}$ & NA & 100\\
            Number of non-target individuals for training & $N^{T'}_{\text{train}}$ & NA & 1000\\
            Number of images of target individual for validation & $N^T_{\text{val}}$ & NA & 20\\
            Number of non-target individuals for validation & $N^{T'}_{\text{val}}$ & NA & 200\\
            Weight to combine losses for two tasks & $w$ & NA & 0.03\\
            \hline
        \end{tabular}
    \end{center}
    \vspace{-1em}
\end{table}

\section{VGGFace2 dataset cleaning}\label{apd:vggface2-cleaning-algorithms}

For images $X^I$ belonging to individual $I$, algorithm~\ref{alg:mislabel-detection} details the procedure for detecting mislabelled images and images without faces in $X^I$. While algorithm~\ref{alg:dup-image-det} details the procedure for detecting duplicates in $X^I$.

\begin{center}
    \begin{algorithm}[!ht]
        \caption{Mislabelled image detection algorithm}\label{alg:mislabel-detection}
        \begin{algorithmic}[1]
            \Inputs{
                $X_I$: Images of individual $I$\\
                $F$: Face detection and recognition algorithm\\
                $T_{mis}$: Distance threshold for flagging mislabelled images\\
            }
            \Output{
                $E$: Mislabelled and faceless images\\
            }
            \Initialize{
                $E \gets $ []\\
                $base \gets \textbf{0}$\\
                $n \gets 0$\\
                $d_{cos} \gets $ Cosine distance function\\
            }

            \LineComment{Step 1: Evaluate base embedding}
            \For{$X_i$ in $X_I$}
            \State{$f \gets F(X_i)$}
            \If{$|f|$ == 1}\Comment{Only 1 face is detected}
            \State{$base \gets base + \frac{f[0]}{||f[0]||_2}$}
            \State{$n \gets n + 1$}
            \EndIf
            \If{$|f|$ == 0}\Comment{No face is detected}
            \State{$E.push(X_i)$}
            \EndIf
            \EndFor
            \State{$base \gets \frac{base}{n}$}

            \LineComment{Step 2: Detect mislabelled images}
            \State{$X_I \gets \{X| X \in X_I, X \notin E\}$}
            \For{$X_i \in X_I$}
            \State{$f \gets F(X_i)$}
            \State{$d = min(d_{cos}(f_j, base) \;\forall\; f_j \in f$)}
            \If{$d > T_{mis}$}
            \State{$E.push(X_i)$}
            \EndIf
            \EndFor
        \end{algorithmic}
    \end{algorithm}
    \vspace{-1em}
\end{center}

\begin{table}[!htbp]
    \caption{False positives per million when matching 10\% DISC21 reference set $R$ to remaining 90\% of $R$}\label{tab:fpm-isc-experiment}
    \vspace{-1em}
    \begin{center}
        \begin{tabular}{ccc}
            \toprule
            \textbf{Models ($H$)} & \textbf{False positives} & \textbf{Estimate of truly false positives}\\
            \midrule
            Single-purpose & 35175 $\pm$ 1475 & 6859 $\pm$ 288\\
            \midrule
            Dual-purpose & 34465 $\pm$ 2233 & 6721 $\pm$ 435\\
            \bottomrule
        \end{tabular}
    \end{center}
    \vspace{-0.5em}
\end{table}

\section{VGGFace2 dataset analysis}\label{apd:vggface2-analysis}

\begin{figure}[!ht]
    \centering
    \includegraphics[width=\linewidth]{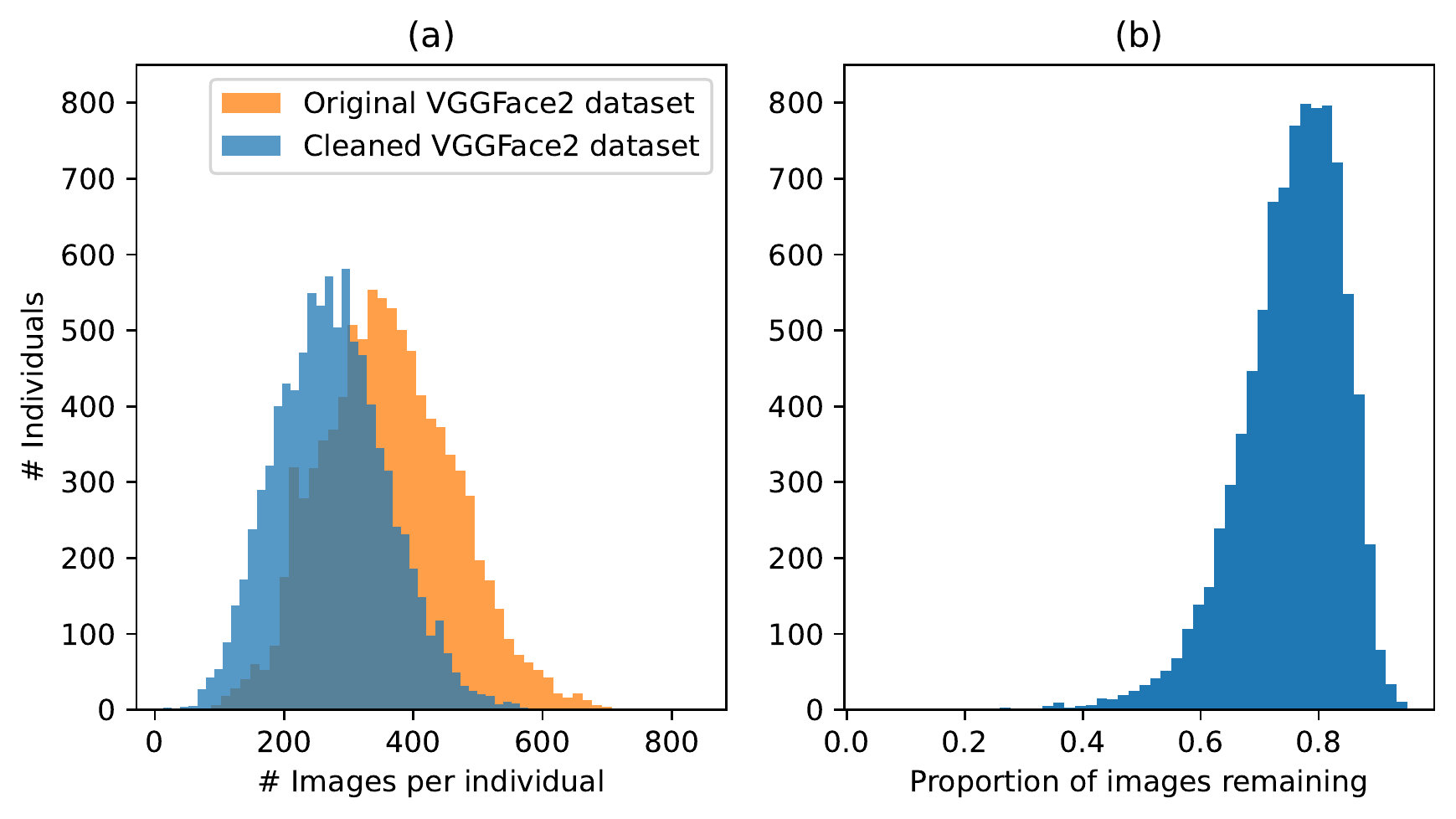}
    \caption{Impact of dataset cleaning on VGGFace2 dataset. a) Distribution of number of images per individual before and after dataset cleaning. b) Distribution of proportion of images remaining per individual.}
    \label{fig:vggface-dataset-analysis}
\end{figure}

Figure~\ref{fig:fr-dataset-images} shows images from VGGFace2~\cite{cao2018vggface2} dataset are more realistic than other recent large-scale facial recognition datasets like WebFace260m~\cite{zhu2021webface260m}. Fig.~\ref{fig:vggface-dataset-analysis} shows the distribution of images per individual in VGGFace2 dataset before and after the dataset cleaning.

\section{Facial recognition performance}

\begin{figure*}[!ht]
    \centering
    \includegraphics[width=\linewidth]{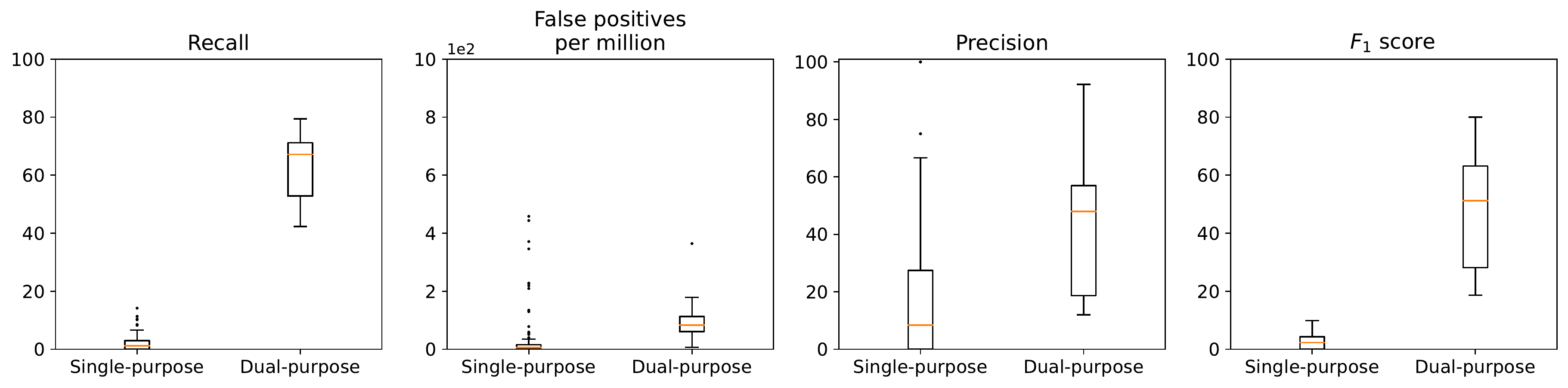}
    \vspace{-1em}
    \caption{Face recognition performance on target individuals of single-purpose and dual-purpose models.}
    \label{fig:fr-target-performance}
\end{figure*}

\begin{figure*}[!ht]
    \centering
    \includegraphics[width=\linewidth]{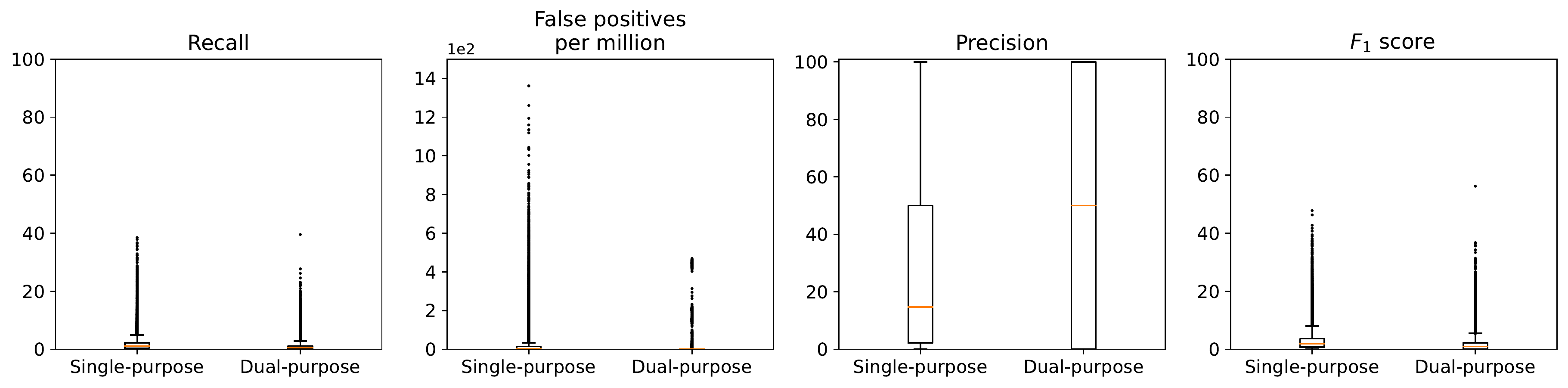}
    \vspace{-1em}
    \caption{Face recognition performance on non-target individuals of single-purpose and dual-purpose models.}
    \label{fig:fr-non-target-performance}
\end{figure*}

Figures~\ref{fig:fr-target-performance} and~\ref{fig:fr-non-target-performance} shows boxplots for the face recognition performance on target and non-target individuals respectively.

\section{Additional evaluations of false positives}\label{apd:fp-evaluation}

Table~\ref{tab:fp-css-evaluation} reports the number of false positives (FP) per million
for a threshold selected with validation precision of 90\%. Table~\ref{tab:fp-css-evaluation-extended} extends the results to the thresholds corresponding to validation precision of 95\% ($T@95$) and 99\% ($T@99$).
As expected, the number of false positives decreases as the detection threshold decreases.

\begin{figure}[!htbp]
    \centering
    \includegraphics[width=\linewidth]{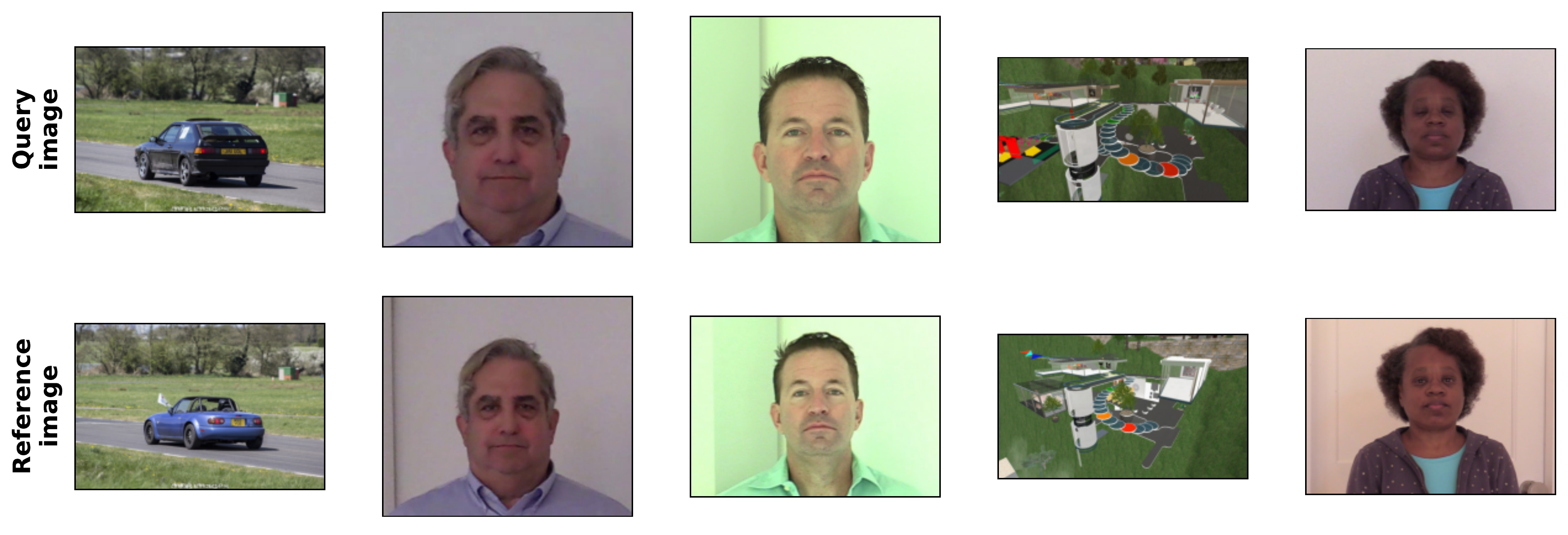}
    \caption{Examples of image pairs flagged as false positive matches when matching 10\% of the DISC21 reference set $R^{\text{primary}}$ against the remaining 90\% of the reference set $R^{\text{primary}}$. The first pair (from left) is a truly false positive. Others are duplicate images.}
    \label{fig:fp-analysis-duplicates}
    \vspace{-1em}
\end{figure}

\begin{table}[!ht]
    \caption{False-positives flagged in client-side scanning setup with different thresholds}\label{tab:fp-css-evaluation-extended}
    \vspace{-1em}
    \begin{center}
        \begin{tabular}{cccc}
            \toprule
            \textbf{Threshold} & {\textbf{Models ($H$)}} & \textbf{Reference set $R$} & \textbf{Reference set $R_d$}\\
            \midrule
            \multirow{2} * {$T@95$} & Single-purpose &  1374.4 $\pm$ 113.2 & 1374.4 $\pm$ 130.7\\
            \cmidrule{2-4}
            & Dual-purpose & 1395.7 $\pm$ 279.0 & 1395.8 $\pm$ 279.0\\
            \midrule
            \multirow{2} * {$T@99$} & Single-purpose & 713.4 $\pm$ 41.1 & 713.4 $\pm$ 43.3\\
            \cmidrule{2-4}
            & Dual-purpose & 702.2 $\pm$ 53.6 & 702.6 $\pm$ 54.2\\
            \bottomrule
        \end{tabular}
    \end{center}
    \vspace{-0.5em}
\end{table}

To understand how the number of FPs is impacted by the datasets being matched, we perform an additional analysis where we match 10\% of the DISC21 reference set $R^{\text{primary}}$ (or 100k images) against the remaining 90\% of the reference set $R^{\text{primary}}$ (containing 900k images) using the default threshold $T$ computed for validation precision of 90\%. The first column in Table~\ref{tab:fpm-isc-experiment} shows that in this setup the FPs flagged are relatively much larger compared to results in Table~\ref{tab:fp-css-evaluation}. However, upon manual inspection of 100 samples of FPs for both single-purpose and dual-purpose models, we observe that only 19.5\% of the matched pair of images are actually false matches while the rest of the pairs were duplicate images as defined in Douze et al.~\cite{douze20212021}. Fig.~\ref{fig:fp-analysis-duplicates} shows examples of some flagged image pairs. We thus estimate the number of \textit{truly false positives} (TFP) by multiplying the observed count of false positives by 0.195. The second column of Table~\ref{tab:fpm-isc-experiment} reports our estimate of TFPs when matching images from the same distribution as the DISC21 reference set $R^{\text{primary}}$. The value is, as expected as the images now come from the same distribution, larger but comparable with the number of FPs observed when matching Imagenet against $R^{\text{primary}}$. We evaluate the estimate of TFPs, similarly as in Table~\ref{tab:fpm-isc-experiment}, for the additional thresholds of $T@95$ and $T@99$. We observe that for $T@99$ number of FPs in this setup are similar to the ones observed in Table~\ref{tab:fp-css-evaluation-extended}.

\begin{algorithm}[!ht]
    \begin{center}
        \caption{Duplicate image detection algorithm}\label{alg:dup-image-det}
        \begin{algorithmic}[1]
            \Inputs{
                $M_I = \{X_1,..,X_n\}$: Images of individual $I$ after\\
                filtering mislabelled and faceless images\\
                $S$: SSCD algorithm\\
                $T_{dup}$: Distance threshold for flagging duplicate\\
                images
            }

            \Output{
                $E$: List of duplicate images to exclude
            }
            \Initialize{
                $E \gets []$\\
                $n \gets |M_I|$\\
                $d_2 \gets$ Euclidean distance function
            }

            \State{$M_I = \text{sorted}(M_I)$}\Comment{Sort image list.}
            \For{$i \in \{1, ..., n\}$}
            \If{$X_i \notin E$}
            \For{$j \in \{i+1,...,n\}$}
            \If{$d_2(S(X_i), S(X_j)) < T_{dup}$}
            \State{$E.push(X_j)$}
            \EndIf
            \EndFor
            \EndIf
            \EndFor
        \end{algorithmic}
    \end{center}
\end{algorithm}

\section{Visualizing activations}\label{apd:activations-vis}

An auditor could inspect the intermediate activations of the model to identify any unusual behavior, e.g., to see if a significant proportion of internal activations are zero for a large sample of images or if images from primary dataset distribution and secondary dataset distribution would result in different parts of the network being activated.

We analyze the output activations of the second to last layer of a dual-purpose model.
Fig.~\ref{fig:activations}A-B shows that for a sample of images without the target individual, most of the neurons are activated as their output activations are, on average, non-zero. 
It also shows that the model behaves similarly for (A) randomly sampled 1,000 query images for the primary task and (B) randomly sampled 1,000 query images of non-target individuals, as shown by the output activations. These results strengthen our claim of a hidden secondary purpose.
For completeness, we also look at the average activations for all the (C) test images of the target individual. 
As expected, we observe a difference, but a realistic auditor will not have access to the images of target individual and thus will not observe the difference.

\begin{table}[!h]
    \caption{Estimate of truly false positives per million when matching 10\% DISC21 reference set $R$ to remaining 90\% of $R$ for different thresholds}\label{tab:fpm-thresholds}
    \begin{center}
        \begin{tabular}{ccc}
            \toprule
            \multirow{3}{*}{\textbf{Models ($H$)}} & \multicolumn{2}{c}{\textbf{Thresholds ($T$}}\\
            \cmidrule{2-3}
            & \textbf{$T@95$} & \textbf{$T@99$}\\
            \midrule
            Single-purpose models $H_s$ & 3145 $\pm$ 139 & 653 $\pm$ 130\\
            \midrule
            Dual-purpose models $H_d$ & 3313 $\pm$ 398 & 720 $\pm$ 47\\
            \bottomrule
        \end{tabular}
    \end{center}
\end{table}

\end{document}